\documentclass[prd,12pt,nofootinbib,preprint]{revtex4}

\usepackage[T1]{fontenc}
\usepackage{amsmath,amssymb}
\usepackage{epsfig}
\usepackage{url}
\usepackage{subfigure}
\usepackage{graphicx}
\usepackage{grffile}
\usepackage[usenames,dvipsnames]{color}
\usepackage{slashed}
\usepackage{array}
\usepackage{multirow}
\usepackage[colorlinks,citecolor=blue]{hyperref}
\usepackage{color}
\usepackage{cancel}
\usepackage{gensymb}

\def\a{\alpha}

\def\b{\beta}

\def\l{\lambda}
\def\m{\mu}

\def\D{\Delta}

\def\q2 {q^2}

\def \met{\not \! E_T }

\def\a {\alpha}
\def\b {\beta}

\def\l {\lambda}

\def\bar {\overline}

\def\be {\begin{equation}}
\def\ee {\end{equation}}
\def\beq {\begin{equation}}
\def\eeq {\end{equation}}

\newcommand{\besub}{\begin{subequations}}
\newcommand{\eesub}{\end{subequations}}
\newcommand{\bea}{\begin{eqnarray}}
\newcommand{\eea}{\end{eqnarray}}

\def\beq{\begin{equation}}
\def\eeq{\end{equation}}
\def\barr{\begin{array}}
\def\earr{\end{array}}

\begin{document}
\title{Doubly charged scalars and vector-like leptons confronting the muon g-2 anomaly and Higgs vacuum stability}

\author{Nabarun Chakrabarty}
\email{nabarunc@iitk.ac.in, chakrabartynabarun@gmail.com}
\affiliation{Centre for High Energy Physics, Indian Institute of Science, C.V.Raman Avenue,
Bangalore 560012, India}
\affiliation{Department of Physics, Indian Institute of Technology Kanpur,
Kanpur, Uttar Pradesh 208016, India}

\begin{abstract} 
The present work introduces new scalar and fermionic degrees of freedom to the Standard Model. While the scalar sector is augmented by a complex scalar triplet and a doubly charged scalar singlet, the fermionic sector is extended by two copies of vector-like leptons. Of these, one copy is an $SU(2)_L$ singlet while the other, an $SU(2)_L$ doublet. We explain how this combination 
can offer a solution to the muon g-2 anomaly and also lead to non-zero neutrino masses. In addition, it is also shown that the parameter regions compliant with the two aforementioned issues can stabilise the electroweak vacuum till the Planck scale, something not possible within the Standard Model alone.    
\end{abstract} 
\maketitle

\section{Introduction}\label{intro}

The discovery of the Higgs boson of mass 125 GeV~\cite{Chatrchyan:2012xdj,Aad:2012tfa} at the Large Hadron Collider (LHC)
completes the particle spectrum of the Standard Model (SM). 
Moreover, the interactions of the boson with SM fermions and gauge bosons are increasingly in agreement with the corresponding SM values. Despite this success, certain pressing inconsistencies within the SM on both  theoretical and experimental fronts continue to vouch for beyond-the-SM (BSM) dynamics. That the SM alone cannot stabilise the electroweak (EW) vacuum up to the Planck scale is one such theoretical shortcoming. More specifically, the SM Higgs
quartic coupling turns negative during renormalisation group (RG) evolution thereby destabilising
the vacuum and the energy scale where that happens can vary several orders of magnitude depending
upon the t-quark mass chosen~\cite{Degrassi:2012ry,Buttazzo:2013uya,Zoller:2014cka,EliasMiro:2011aa,Isidori:2001bm}. However, additional bosonic degrees of freedom over and above the SM ones can potentially offset this destabilising effect coming from the t-quark (see the references in \cite{Swiezewska:2016rrp}). This calls for extending the scalar sector of the SM. 

One crucial shortcoming of the SM on the experimental side
is its inability to predict non-zero neutrino masses.
In addition, the longstanding deviation in the experimentally measured value of the muon anomalous magnetic moment from
its SM prediction also necessitates BSM dynamics. A 3.7$\sigma$
discrepancy exists between theoretical calculations within the SM and experimental data,
quoting~\cite{Bennett:2006fi,Miller:2007kk,doi:10.1146/annurev-nucl-031312-120340,Keshavarzi:2018mgv,Blum:2018mom,Aoyama:2020ynm}
\bea
\Delta a_\mu = (2.706 \pm 0.726) \times 10^{-9}.
\eea
This deviation is seen as an evidence of the presence of BSM dynamics. One must note that the discrepancy will be put to further tests at the FNAL \cite{Grange:2015fou}
and J-PARC \cite{Iinuma:2011zz} experiments in the near future. 

Appropriately augmenting the SM by additional fields can lead to a non-zero neutrino mass via the seesaw mechanism.
Of these, the popular Type-II seesaw~\cite{PhysRevD.22.2227,Magg:1980ut,Lazarides:1980nt} employs a complex scalar $SU(2)_L$ triplet and is also known to be attractive from the perspective of baryogenesis and collider sigantures. In fact, it has also been shown to address the vacuum instability problem~\cite{Chun:2012jw,Dev:2013ff,Chakraborty:2014xqa}. However, despite such enticing aspects, the Type-II seesaw model is known to generate a negative contribution to the muon g-2~\cite{Fukuyama:2009xk}, and hence, cannot account for the observed discrepancy. And this can be attributed to the completely left-chiral Yukawa interactions of the scalar triplet.

New Physics (NP) models comprising vector-like leptons (VLLs) 
have interesting phenomenological implications. Having novel origins such as Grand Unification~\cite{Thomas:1998wy,Freitas:2020ttd}, 
the SM suitably
augmented by VLLs can in fact explain the muon g-2 anomaly~\cite{Kannike:2011ng,Dermisek:2013gta,Megias:2017dzd,Crivellin:2018qmi}. However, the minimal VLL scenario does not offer solutions to the neutrino mass and vacuum instability problems, Moreover, it gets rather constrained by the measurements of the Higgs to dimuon decay made by ATLAS~\cite{Aad:2020xfq} and CMS~\cite{Sirunyan:2018hbu}. Some recent solutions to the muon anomaly involving together vector leptons and additional scalar multiplets can be seen in~\cite{Frank:2020smf,Chun:2020uzw,Chen:2020tfr,Jana:2020joi,deJesus:2020upp}.

In this work, we extend the Type-II seesaw model by an doubly charged $SU(2)_L$ singlet scalar and VLLs. A doubly charged scalar is an ingredient of certain classes of NP models, the minimal left-right symmetric model (LRSM) augmented with scalar triplets being an example.  That is, the triplets $\Delta_L$ (\textbf{1,3,1},2) and $\Delta_R$ (\textbf{1,1,3},2) are introduced under the LRSM gauge group $SU(3)_c \times SU(2)_L \times SU(2)_R \times U(1)_{B-L}$ \cite{Iso:2009nw}, over and above the minimal field content. On the other hand, some investigations involving a scalar triplet and VLLs are~\cite{Bahrami:2013bsa,Bahrami:2015mwa,Bahrami:2016has}. We thus have two doubly charged scalars in this scenario instead of one as in the case of ordinary Type-II seesaw\footnote{\cite{Chakrabarty:2018qtt} presents explanations the muon anomaly in models featuring a two doubly charged scalars but no additional fermions over and above the SM ones.}. The VLLs include both doublets and singlets under $SU(2)_L$, the latter carrying one unit of electric charge.
We explain how a positive contribution of the requisite
magnitude to the muon g-2 can be obtained in this framework by virtue of a non-zero mixing of the two doubly charged bosons. We also demonstrate that tuning the Yukawa interactions and the triplet vacuum expectation value (VEV) correctly can help evade the constraints coming from
the non-observation of charged lepton flavour violation (CLFV)~\cite{Calibbi:2017uvl}. In addition, we compute the one-loop RG equations corresponding to this model and subsequently show that the 
combined results of neutrino mass, muon g-2 and LFV comply with a stable EW vacuum till the Planck scale.

This paper is organised as follows. We introduce the theoretical framework in section \ref{model} and list the various constraints in section \ref{constraints}. Section \ref{g-2} demonstrates the role of chirality-flip in generating a positive contribution to muon g-2 while predicting non-zero neutrino masses and suppressing CLFV. Section \ref{vacstab} presents an analysis combining vacuum stability, muon g-2 and the various relevant constraints. We summarise in section \ref{summary}. Various important formulae are relegated to the Appendix. 

\section{Model description
}\label{model}

In this model, the scalar sector of the SM is augmented by an $SU(2)_L$ complex scalar triplet $\Delta$ and a doubly charged scalar singlet $k^{++}$. In addition, the following VLL multiplets are also included:
\besub
\bea
L_{L,R} = 
\begin{pmatrix}
N_{L,R} \\
E_{L,R}
\end{pmatrix};~~~~
E^\prime_{L,R}.
\eea
\eesub
The quantum numbers of the various relevant fields are shown in Table~\ref{bsm}.

\begin{table}
\centering
\begin{tabular}{ |c|c| } 
\hline
Field & $SU(3)_c \times SU(2)_L \times U(1)_Y$ \\ 
\hline \hline 
$\Delta$ & $(\mathbf{1,3},1)$ \\ \hline 
$k^{++}$ & $(\mathbf{1,1},2)$\\ \hline 
$L_{L,R}$ & $(\mathbf{1,2},-1/2)$\\ \hline 
$E^\prime_{L,R}$ & $(\mathbf{1,1},-1)$\\ \hline 
\end{tabular}
\caption{Quantum numbers of the relevant fields under the SM gauge group.}
\label{bsm}
\end{table}

As for how the additional fields interact, we first show the scalar potential below:
\bea
V &=& V_2 + V_3 + V_4,
\eea
where $V_n$ for $n$ = 2,3,4 describes dimension-$n$ scalar operators. Thus,
\besub
\bea
V_2 &=& \m^2_{\phi} (\phi^{\dagger} \phi) + M^2_{\Delta} \text{Tr} (\Delta^{\dagger}\Delta) + M^2_S |k^{++}|^2, 
\\
V_3 &=& \mu_1^{}\, \phi^T (i \sigma_2^{}) \Delta^{\dagger} \phi 
 + \mu_2^{}\, \text{Tr}\big( \Delta^{\dagger} \Delta^{\dagger}\big) k^{++} + \text{h.c.} \\
V_4 &=& \frac{\l}{2} (\phi^{\dagger} \phi)^2
+ \frac{\l_1}{2} [\text{Tr} (\Delta^{\dagger}\Delta)]^2
+ \frac{\l_2}{2}\Big([\text{Tr} (\Delta^{\dagger}\Delta)]^2 - \text{Tr}(\Delta^{\dagger}\Delta\Delta^{\dagger}\Delta) \Big)  + \frac{\l_3}{2} |k^{++}|^4 \nonumber \\
&&
+ \l_4 \phi^{\dagger} \phi \text{Tr} 
(\Delta^{\dagger}\Delta)
+ \l_5 \phi^{\dagger} \big[\Delta, \Delta^\dagger \big] \phi + \l_6 \phi^{\dagger}\phi |k^{++}|^2
+ \l_7 \text{Tr}(\Delta^{\dagger}\Delta) |k^{++}|^2 \nonumber \\
&&
+ \l_8^{} \big({\tilde{\phi}}^{\dagger} \Delta \phi k^{--} +  \text{h.c.}\big). 
\eea 
\eesub
We choose all parameters in the scalar potential to be real to annul CP-violation. To state the obvious, the scalar interactions involving $k^{++}$ are the additional ones \emph{w.r.t.} the ordinary Type-II case. The scalar doublet and the triplet can be parameterised as under.
\besub
\bea
\phi = 
\begin{pmatrix}
    \phi^+ \\
    \frac{1}{\sqrt{2}}(v + \phi_0 + i \eta_0)
  \end{pmatrix},~~~~~
\Delta = 
\begin{pmatrix}
   \frac{\delta^+}{\sqrt{2}} & \delta^{++} \\
    \frac{1}{\sqrt{2}}(v_\D + \delta_0 + i \chi_0) & 
    -\frac{\delta^+}{\sqrt{2}}.
  \end{pmatrix}  
\eea
\eesub
Here, $v$ and $v_\Delta$ denote the VEVs acquired by the 
CP-even neutral components of $\phi$ and $\Delta$ respectively. The scalar potential leads to the following mixings in the CP-even, CP-odd and singly charged sectors. 
\besub
\bea
\begin{pmatrix}
    \phi_0 \\
    \delta_0
  \end{pmatrix} =
\begin{pmatrix}
    \text{cos}\a & \text{sin}\a \\
    -\text{sin}\a & \text{cos}\a
  \end{pmatrix}
\begin{pmatrix}
    h \\
    H
  \end{pmatrix} \\
\begin{pmatrix}
    \eta_0 \\
    \chi_0
  \end{pmatrix} =
\begin{pmatrix}
    \text{cos}\b & \text{sin}\b \\
    -\text{sin}\b & \text{cos}\b
  \end{pmatrix}
\begin{pmatrix}
    G^0 \\
    A
  \end{pmatrix} \\  
\begin{pmatrix}
    \phi^+ \\
    \delta^+
  \end{pmatrix} =
\begin{pmatrix}
    \text{cos}\gamma & \text{sin}\gamma \\
    -\text{sin}\gamma & \text{cos}\gamma
  \end{pmatrix}
\begin{pmatrix}
    G^+ \\
    H^+
  \end{pmatrix}
\eea 
\eesub 
We note that the aforementioned mixings are identical to what happens in the ordinary Type-II case. The mixing angles $\a,\b,\gamma$ are determined to be
\besub
\bea
\text{tan} \a &=& -\frac{4 v_\D}{v} \frac{M^2_\D + 
\frac{1}{2} \l_1 v^2_\D}{M^2_\D + \frac{3}{2}
\l_1 v^2_\D + \frac{1}{2}(\l_4 - \l_5 - 2\l)v^2}, \\
\text{tan} \b &=& -\frac{2 v_\D}{v},\\
\text{tan} \gamma &=& -\frac{\sqrt{2} v_\D}{v}.
\eea
\eesub 
We choose to adopt the $v_\D << v$ limit throughout wherein the expressions for the physical masses simplify to
\besub
\bea
M_h^2 &\simeq& \l v^2, \\
M_H^2 &=& M_A^2 \simeq M^2_\D + \frac{1}{2}(\l_4 + \l_5)v^2, \\
M^2_{H^+} &=&  M^2_\D + \frac{1}{2}\l_4 v^2.
\eea
\eesub
In addition to the above, the doubly charged scalars also 
mix for $\l_8 \neq 0$. The mass terms have the following form for $v_\Delta << v$ and $\mu_2 << \l_8 v$:
\besub
\bea
\mathcal{L}^{++}_m &=& 
\begin{pmatrix}
\delta^{--} & k^{--}
\end{pmatrix}
\begin{pmatrix}
M^2_{\Delta} + \frac{1}{2} \l_4^{} v^2 & \frac{1}{2}\l_8 v^2 \\
\frac{1}{2}\l_8 v^2 & M^2_S
 + \frac{1}{2} \l_6 v^2
\end{pmatrix}
\begin{pmatrix}
\delta^{++} \\
 k^{++}
\end{pmatrix}\label{matrix_doublycharged}
\eea
\eesub 

Diagonalising Eq.(\ref{matrix_doublycharged}) by rotating ($\delta^{++},k^{++}$) by an angle $\theta$ leads to the mass eigenstates $H_{1,2}^{++}$ with masses 
$M^{++}_{1,2}$. Thus,
\besub
\bea
\begin{pmatrix}
    \delta^{++} \\
     k^{++}
  \end{pmatrix} =
\begin{pmatrix}
    \text{cos}\theta & \text{sin}\theta \\
    -\text{sin}\theta & \text{cos}\theta
  \end{pmatrix}
\begin{pmatrix}
    H_1^{++} \\
    H_2^{++}
  \end{pmatrix}
\eea
\eesub

We also list below the expressions for the masses of $H^{++}_{1,2}$ and $\theta$ for $v_\Delta < < v$:
\besub
\bea
(M_{1,2}^{++})^2 &=& \frac{1}{2}\big[(A + B) \pm 
\sqrt{(A - B)^2 + 4 C^2}\big]\label{mpp} ~, \\
\text{tan} 2\theta &=& \frac{2 C}{B - A} ~, ~~~\text{where}\label{tan} \\
A &=& M^2_{\Delta} + \frac{1}{2} \l_4 v^2 ~,\label{Aprime} \\
B &=& M^2_S + \frac{1}{2} \l_6^{} v^2 ~,\label{Bprime} \\
C &=& \frac{1}{2}\l_8^{} v^2 ~.\label{Cprime} 
\eea
\eesub

We now come to discussing the fermionic interactions. First, the bare mass terms of the VLLs and their interactions with the Higgs doublet $\phi$ read
\besub
\bea
\mathcal{L}^{\text{VLL}}_{Y,\phi} &=& -M \bar{L_L} L_R - M^\prime \bar{E_L^\prime} E_R^\prime - y_4 \bar{L}_L \phi E_R^\prime 
- y_4^\prime \bar{L}_R \phi E_L^\prime + \text{h.c.}
\eea
\eesub
We neglect here the mixings of the VLLs with the SM leptons
for simplicity\footnote{The mixings, even if allowed, are rendered small from the non-observation of CLFV. This has been explicitly demonstrated in \cite{Ishiwata:2013gma} for VLLs having quantum numbers identical to the present scenario. Therefore, they anyway do not majorly modify the muon g-2 prediction in this model thereby justifying the choice. Other constraints on such mixings, although subleading to CLFV, stem from the measurement of $p p \to h \to \mu \mu$ \cite{Dermisek:2013gta} and $p p \to h \to 4 l$ \cite{Dermisek:2014cia}.}. The mass terms of the VLLs then take the form
\besub
\bea
\mathcal{L}^{\text{VLL}}_{Y,\phi} \supset
-\begin{pmatrix}
 \bar{E_R} & \bar{E^\prime_R}
 \end{pmatrix}
\begin{pmatrix}
    M & \frac{y^\prime_4 v}{\sqrt{2}} \\
    \frac{y_4 v}{\sqrt{2}} & M^\prime
\end{pmatrix}  
\begin{pmatrix}
E_L \\
E^\prime_L
 \end{pmatrix} + \text{h.c.} \label{VLL_mat}
\eea
\eesub
The non-hermitian matrix in Eq.(\ref{VLL_mat}) is diagonalised by a bi-unitary transformation of the form
\bea
{U_R}^\dagger M_V U_L &=& M_V^d,
\eea
where 
\bea
M_V = 
\begin{pmatrix}
    M & \frac{y^\prime_4 v}{\sqrt{2}} \\
    \frac{y_4 v}{\sqrt{2}} & M^\prime
\end{pmatrix}, ~~~
M_V^d = 
\begin{pmatrix}
    M_1 & 0 \\
    0 & M_2
\end{pmatrix}, ~~~
U_{L(R)} = 
\begin{pmatrix}
    \text{cos}\a_{L(R)} & \text{sin}\a_{L(R)} \\
    -\text{sin}\a_{L(R)} & \text{cos}\a_{L(R)}
  \end{pmatrix}.
\eea
Therefore, the VLLs in the mass basis, i.e.,  
$E_{L(R)_1}$ and $E_{L(R)_2}$, are obtained by rotating the flavour basis as 
\besub
\bea
\begin{pmatrix}
    E_{L(R)} \\
    E_{L(R)}^\prime
  \end{pmatrix} =
U_{L(R)}
\begin{pmatrix}
    E_{{L(R)}_1} \\
    E_{{L(R)}_2}
  \end{pmatrix}.
\eea
\eesub

Next, denoting an SM lepton doublet (singlet) as $L_{\a L}$ 
($l_{\a R}$), Yukawa interactions with the triplet $\Delta$ can be written as
\besub
\bea
\mathcal{L}_{Y,\Delta} &=& \mathcal{L}^{\text{SM}}_{Y,\Delta} + \mathcal{L}^\text{VLL}_{Y,\Delta}, \\
\mathcal{L}^{\text{SM}}_{Y,\Delta} &=& -\sum_{\a,\b=e,\mu,\tau}
y_{\Delta}^{\a \b} \bar{L^c_{\a L}}~i\sigma_2 \Delta~L_{\b L} + \text{h.c.}, \\
\mathcal{L}^\text{VLL}_{Y,\Delta} &=& -2\sum_{\a =e,\mu,\tau}
y_{\Delta}^{\a 4} \bar{L^c_{\a L}}~i\sigma_2 \Delta~L_L 
-y_{\Delta}^{4 4} \bar{L^c_{L}}~i\sigma_2 \Delta~L_L
+ \text{h.c.}
\eea
\eesub
One notes that the term $\mathcal{L}^{\text{SM}}_{Y,\Delta}$
parameterises the interactions involving the SM leptons and $\Delta$ and is also present in the minimal Type-II model. On the other hand, $\mathcal{L}^{\text{VLL}}_{Y,\Delta}$ describes how the VLLs interact with $\Delta$
and is an addition over the minimal Type-II. Finally, we describe the Yukawa interactions involving $k^{++}$ below.
\besub
\bea
\mathcal{L}_{Y,k^{++}} &=& \mathcal{L}^{\text{SM}}_{Y,k^{++}} + \mathcal{L}^\text{VLL}_{Y,k^{++}}, \\
\mathcal{L}^{\text{SM}}_{Y,k^{++}} &=&
-\sum_{\a,\b=e,\mu,\tau}
y_{S}^{\a \b}~\bar{l^c_{\a R}}~l_{\b R} k^{++}
 + \text{h.c.}, \\
\mathcal{L}^{\text{VLL}}_{Y,k^{++}} &=&
-2\sum_{\a=e,\mu,\tau} 
y_S^{\a 4}~\bar{l^c_{\a R}}~E_R^\prime k^{++}
- y_S^{4 4}~\bar{{E^\prime}^c_{R}}~E_R^\prime k^{++}
 + \text{h.c.} 
\eea
\eesub

It is convenient to describe the present framework in terms of masses and mixing angles. The following scalar quartic couplings can be solved in terms of physical scalar masses and the mixing angle $\theta$ as under.
\besub
\bea
\l &=& \frac{M^2_h}{v^2}, \\
\l_4 &=& \frac{2(M^2_{H^+} - M^2_\Delta)}{v^2}, \\
\l_5 &=& \frac{2(M^2_H - M^2_{H^+})}{v^2}, \\
\l_6 &=& \frac{2\big[ (M_1^{++})^2 \text{sin}^2\theta + 
(M_2^{++})^2 \text{cos}^2\theta - M_S^2 \big]}{v^2}, \\
\l_8 &=& \frac{2 \text{sin}\theta \text{cos}\theta
\big[ (M_2^{++})^2 - (M_1^{++})^2 \big]}{v^2}\label{l8}. 
\eea
\eesub
The independent parameters in the scalar sector are therefore $\{v,v_\Delta,\mu_2,M_h,M_H,M_H^+,M_1^{++},M_2^{++},M_\Delta,M_S,\l_1,\l_2,\l_3,\l_7\}$. Of these, we fix $M_h$ = 125 GeV and $v \simeq$ 246 GeV for $v_\Delta << v$.

A non-zero $v_\Delta$ leads to non-zero neutrino-mass elements of the form $m_\nu^{\a\b} = \sqrt{2} 
y_\Delta ^{\a\b} v_\Delta$. This necessitates $y_\Delta^{\a\b}$ to be complex. All other Yukawa couplings
are taken real since they do not participate in neutrino mass generation. One can also eliminate $y_4,y_4^\prime$ in favour of the VLL masses and $\a_L,\a_R$ as
\besub
\bea
y_4 &=& \frac{\sqrt2}{v}( M_2 \text{sin}\a_L~\text{cos}\a_R- M_1 \text{cos}\a_L~\text{sin}\a_R  ), \label{y4} \\
y_4^\prime &=& \frac{\sqrt2}{v}( M_2 \text{cos}\a_L~\text{sin}\a_R  - M_1\text{sin}\a_L~\text{cos}\a_R  ). \label{y4p}
\eea
\eesub
The neutral member of the VLL multiplet, $N$, then has
the mass
\bea
M_N = M = M_1 \text{cos} \a_L \text{cos} \a_R + M_2 \text{sin} \a_L \text{sin} \a_R.
\eea
The independent paramaters in the fermionic sector are therefore $\{m_\nu^{\a\b},y_\Delta^{\a 4},y_\Delta^{4 4},y_S^{\a \b},y_S^{\a 4},y_S^{4 4},M_1,M_2,\a_L,\a_R\}$ of which $m_\nu^{\a\b}$ are sharply constrained by the neutrino-oscillation data. 

To this end, one could think of a spin-off scenario sharing a similar field content as the present one. An additional $Y$ = 0
$SU(2)_L$ singlet scalar $S$ (say) can be additionally introduced (see \cite{Das:2020hpd} and the references therein for a discussions on the scalar singlet assisted scotogenic model) and a $\mathbb{Z}_2$ symmetry can be further invoked under which 
$\{S,\Delta,k^{++},L_{L,R},E^\prime_{L,R} \} \to \{-S,-\Delta,-k^{++},-L_{L,R},
-E^\prime_{L,R} \}$ while the SM fields are even. Such a construct has several implications. First, it enforces $V_3,~\mathcal{L}_{Y,\Delta}^{SM},
~\mathcal{L}_{Y,k^{++}}^{SM} \to 0$ and also obviates mixing between the SM leptons and the VLLs. Secondly, the lightest neutral particle in the $\mathbb{Z}_2$-odd sector can be a candidate for dark matter (DM). Thirdly, a non-zero neutrino mass in this case is realised at one-loop with the VLLs and $\mathbb{Z}_2$-odd scalars circulating in the loop. Therefore, the model introduced in this paper can be a precursor to a future study involving DM that would essentially retain the main mechanism responsible for the muon g-2 enhancement as detailed in this study.

\section{Possible constraints} \label{constraints}
We list in this section various constraints on the present
framework from both theory and experiments.

\subsection{Theoretical constraints}

The bounds $|\l_i| \leq 4\pi, |y_i| < \sqrt{4\pi}$ ensure that the theory remains perturbative, where $\l_i$($y_i$) denotes a generic quartic(Yukawa) coupling. 

The following conditions ensure that the scalar potential remains bounded from below (BFB) for large field values of the constituent scalar fields:
\besub
\bea
V_1 \equiv \l > 0, \\
V_2 \equiv \l_1 > 0, \\
V_3 \equiv 2\l_1 + \l_2 > 0, \\
V_4 \equiv \l_3 > 0, \\
V_5 \equiv \l_4 - \l_5 + \sqrt{\l \l_1} > 0, \\
V_6 \equiv \l_4 + \l_5 + \sqrt{\l \l_1} > 0, \\
V_7 \equiv \l_4 + \l_5 + 
V_8 \equiv \sqrt{\l \big(\l_1 + \frac{\l_2}{2} \big)} > 0, \\
V_9 \equiv \l_4 - \l_5 + 
\sqrt{\l \big(\l_1 + \frac{\l_2}{2} \big)} > 0, \\
V_{10} \equiv \l_6 + \sqrt{\l \l_3} > 0, \\
V_{11} \equiv \l_7 + \sqrt{\l_1 \l_3} > 0, \\
V_{12} \equiv \l_7 + \sqrt{\l_3 \big(\l_1 + \frac{\l_2}{2} \big)} > 0.
\eea
\eesub
A given condition in the aforementioned set comes from demanding the 
scalar potential remains BFB in a given direction in the field space.

Additional constraints on the quartic couplings come from unitarity. A tree-level 2 $\to$ 2
scattering matrix can be constructed between various two particle states consisting of charged
and neutral scalars~\cite{PhysRevD.7.3111,PhysRevD.16.1519}. Unitarity demands that the absolute value of each eigenvalue of
the aforementioned  matrix must be bounded from above at 
8$\pi$. The conditions for the present scenario are\footnote{The expressions have been checked with \cite{PhysRevD.84.095005} in the appropriate limit.}
\besub
\bea
|\l_1 + \l_2| \leq 8\pi, \\
|\l_4 - 2\l_5| \leq 8\pi, \\
|\l_4 \pm \l_5| \leq 8\pi, \\
|2 \l_1 + 3\l_2| \leq 16\pi, \\
\Big(\l + \l_1 - \l_2 \pm \sqrt{(\l - \l_1 + \l_2)^2 + 16 \l_5^2} \Big) \leq 16\pi, \\
\Big(\l + \l_7 \pm \sqrt{(\l - \l_7)^2 + 8 \l_8^2} \Big) 
\leq 16\pi, \\
\Big(\l_4 + 2\l_5 + \l_6 \pm \sqrt{(\l_4 - 2\l_5 - \l_6)^2 + 24 \l_8^2} \Big) \leq 16\pi.
\eea
\eesub
In addition, these bounds obtained from demanding perturbativity and
a BFB as well as unitary scalar potential must be imposed at each energy scale while evolving the quartic couplings under RG. 

\subsection{Neutrino mass}
The $U_{\text{PMNS}}$ matrix diagonalizes the neutrino mass matrix $m_\nu$, {\it i.e.},
\besub
\bea
&&
m_\nu = U_{\text{PMNS}}^* ~m_\nu^{\text{diag}} ~U_{\text{PMNS}}^T
~, \label{nu}\\
&& \text{with}
~U_{\text{PMNS}} = V_{\text{PMNS}} \times 
\text{diag}(1,e^{i \a_{21}/2},,e^{i \a_{31}/2}) ~\mbox{and} \\
&& V_{\text{PMNS}} =  \begin{pmatrix}
    c_{12} c_{13} & s_{12} c_{13} &  s_{13} e^{-i \delta_{CP}}\\
 -s_{12}c_{23} - c_{12}s_{23}s_{13} e^{i \delta_{CP}} & c_{12}c_{23} - s_{12}s_{23}s_{13} e^{i \delta_{CP}} & s_{23}c_{13} \\
  s_{12}s_{23}-c_{12}c_{23}s_{13} e^{i \delta_{CP}} 
  & -c_{12}s_{23} - s_{12}c_{23}s_{13} e^{i \delta_{CP}} & c_{23} c_{13}
  \end{pmatrix}
  ~,
\eea
\eesub
where $s_{ij} =\sin\theta_{ij}$, $c_{ij} =\cos\theta_{ij}$, $\delta_{CP}^{}$ is the Dirac phase, and $\a_{21}^{}$ and $\a_{31}^{}$ are the Majorana phases. We  fix the neutrino oscillation parameters to their central values~\cite{Patrignani:2016xqp} as
\bea
&&
\text{sin}^2\theta_{12} = 0.307 ~,~ 
~\text{sin}^2\theta_{23} = 0.510 ~,~
~\text{sin}^2\theta_{13} = 0.021 ~, \nonumber \\
&&
\Delta m^2_{21}
 = 7.45 \times 10^{-5} ~\text{GeV}^2 ~,
~\Delta m^2_{32} = 2.53 \times 10^{-3} ~\text{GeV}^2 ~,  \nonumber \\
&&
\delta_{CP} =  1.41\pi ~,~ 
~\a_{21} = \a_{31} = 0 ~. \label{nuparam_fixed}
\eea
The mass of the lightest neutrino and Majorana phases are assumed to vanish in
the present analysis.

\subsection{Collider limits on VLL masses}

Limits on the VLL masses are weak for negligible mixing of the VLLs with the SM leptons which is the case here.
A limit in case of an heavy charged lepton from colliders reads $M,M^\prime > 102.6$ GeV~\cite{PhysRevD.98.030001}. A weak limit $\sim \mathcal{O}$(MeV) on masses neutral leptons comes from Big Bang Nucleosynthesis (BBN)~\cite{PhysRevD.98.030001}. We therefore take $M_N,M_1,M_2 > 110$ GeV for the subsequent analysis.

\subsection{$T$-parameter}
We derive the contribution of the VLLs to the electroweak $T$-parameter~\cite{PhysRevD.46.381} following \cite{PhysRevD.96.015006,Frank:2020smf}.
\bea
\Delta T_{\text{VLL}} &=& \frac{1}{4\pi s^2_w c^2_w} 
\Big(2 h_+(r_1,r_N) + (\text{sin}^2\a_L + \text{sin}^2\a_R) 
[-h_+(r_1,r_N) + h_+(r_2,r_N)] \nonumber \\
&&
+ 2 \text{cos}\a_L \text{cos}\a_R~h_-(r_1,r_N)
+ 2 \text{sin}\a_L \text{sin}\a_R~h_-(r_2,r_N) \Big),
\eea
where
\besub
\bea
h_+(x,y) &=& \frac{x + y}{2} - \frac{xy}{x - y}\text{log}
\Big( \frac{x}{y} \Big);~~~~x \neq y, \nonumber \\
&=& 0;~~~~x \neq y. \\
h_-(x,y) &=& \sqrt{xy} \Bigg[\frac{x + y}{x - y} \text{log}
\Big( \frac{x}{y} \Big) -2 \Bigg];~~~~x \neq y, \nonumber \\
&=& 0;~~~~x \neq y.
\eea
\eesub
Also, $r_N = \Big(\frac{M_N}{M_Z}\Big)^2$ and $r_{1,2} = \Big(\frac{M_{1,2}}{M_Z}\Big)^2$. As for any scalar contribution, the $T$-parameter has a counter term at quantum level unlike the SM and its
multi-doublet Higgs extensions. This additional counter term stems
from the renormalisation of $v_\Delta$. In order to fit the
experimental data, potentially large contribution due to the mass
splittings to $T$ can be absorbed by the counterterm. Hence, after renormalisation, we do not expect stringent constraints on scalar mass splittings. The scalar contribution is therefore ignored in this work. Taking the global electroweak fit~\cite{PhysRevD.98.030001}, we impose the limit
$\Delta T_{\text{VLL}} = 0.07 \pm 0.12$ at 2$\sigma$.

\subsection{$h \to \gamma \gamma$ signal strength}
The dominant amplitude for the $h \to \gamma \gamma$ in the SM reads
\besub
\bea
\mathcal{M}^{\text{SM}}_{h \to \gamma \gamma} &=& 
\frac{4}{3} A_{1/2}\Big(\frac{M^2_h}{4 M^2_t}\Big)
+ A_1\Big(\frac{M^2_h}{4 M^2_W}\Big).
\label{htogaga_SM}
\eea
\eesub
We have neglected the small effect of fermions other than the $t$-quark in Eq.(\ref{htogaga_SM}).
The presence of additional charged scalars and leptons implies that additional one-loop contributions to the $h\gamma\gamma$ amplitude shall arise thereby modifying  
the corresponding decay width \emph{w.r.t.} the SM. The amplitude stemming from the charged scalars $H^+,H_{1,2}^{++}$ reads~\cite{Djouadi:2005gi,Djouadi:2005gj,Arhrib:2011vc}
\besub
\bea
\mathcal{M}^{\text{CS}}_{h \to \gamma \gamma} &=& 
\frac{\l_{h H^+ H^-} v}{2 M^2_{H^+}} A_0\bigg(\frac{M^2_h}{4 M^2_{H^+}}\bigg)
+ \frac{2\l_{h H_1^{++} H_1^{--}} v}{(M_1^{++})^2} A_0\bigg(\frac{M^2_h}{4 (M_1^{++})^2} \bigg)\nonumber \\
&&
+ \frac{2\l_{h H_2^{++} H_2^{--}} v}{(M_2^{++})^2} A_0\bigg(\frac{M^2_h}{4 (M_2^{++})^2} \bigg),
\label{htogaga_scalar}
\eea
\eesub
Where,
\besub
\bea
\l_{h H^+ H^-} &=& \l_4 v, \\
\l_{h H_1^{++} H_1^{--}} &=& v \big\{(\l_4 - \l_5) 
c^2_\theta + \l_6 s^2_\theta - 2 \l_8 s_\theta c_\theta\big\}, \\
\l_{h H_1^{++} H_2^{--}} &=& v \big\{(\l_4 - \l_5) 
s^2_\theta + \l_6 c^2_\theta + 2 \l_8 s_\theta c_\theta\big\}.
\eea
\eesub
Similarly, the VLLs contribute the following to the amplitude
\besub
\bea
\mathcal{M}^{\text{VLL}}_{h \to \gamma \gamma} &=& 
\sum_{i=1,2} y_{h E_i E_i} A_{1/2}\Big(\frac{M^2_h}{4 M^2_i}\Big).
\label{htogaga_VLL}
\eea
\eesub
with
\besub
\bea
y_{h E_1 E_1} &=& \frac{1}{2 v} \Big[ M_1 
\Big(-1 + \text{cos}(2\alpha_L) \text{cos}(2\alpha_R) \Big) + M_2 \Big( \text{sin}(2\alpha_L) \text{sin}(2\alpha_R) \Big)\Big], \\
y_{h E_2 E_2} &=& \frac{1}{2 v} \Big[ M_2 
\Big(-1 + \text{cos}(2\alpha_L) \text{cos}(2\alpha_R) \Big) + M_1 \Big( \text{sin}(2\alpha_L) \text{sin}(2\alpha_R) \Big)\Big]. 
\eea
\eesub
The total amplitude and the decay width then become
\bea
\mathcal{M}_{h \to \gamma \gamma} &=& 
\mathcal{M}^{\text{SM}}_{h \to \gamma \gamma} +
\mathcal{M}^{\text{CS}}_{h \to \gamma \gamma} +
\mathcal{M}^{\text{VLL}}_{h \to \gamma \gamma}, \\
\Gamma_{h \to \gamma \gamma} &=& \frac{G_F \a_{em}^2 M_h^3}{128 \sqrt{2} \pi^3} |\mathcal{M}_{h \to \gamma \gamma}|^2.\eea
where $G_F$ and $\a_{em}$ denote respectively the Fermi constant and the QED fine-structure constant. The various loop functions are listed below~\cite{Djouadi:2005gj}.
\besub
\bea
A_{1/2}(x) &=& \frac{2}{x^2}\big((x + (x -1)f(x)\big), \\
A_1(x) &=& -\frac{1}{x^2}\big((2 x^2 + 3 x + 3(2 x -1)f(x)\big), \\
A_0(x) &=& -\frac{1}{x^2}\big(x - f(x)\big),  \\
\text{with} ~~f(x) &=& \text{arcsin}^2(\sqrt{x}); ~~~x \leq 1 
\nonumber \\
&&
= -\frac{1}{4}\Bigg[\text{log}\frac{1+\sqrt{1 - x^{-1}}}{1-\sqrt{1 - x^{-1}}} -i\pi\Bigg]^2; ~~~x > 1.
\eea
\eesub
where $A_{1/2}(x), A_1(x)$ and 
$A_0(x)$ are the respective amplitudes for the spin-$\frac{1}{2}$, spin-1 and spin-0 particles in the loop.
The signal strength for the $\gamma\gamma$ channel is defined as
\bea
\mu_{\gamma\gamma} &=&  \frac{\sigma(pp \to h)\text{BR}(h \to \gamma \gamma)}{\Big[\sigma(pp \to h)\text{BR}(h \to \gamma \gamma)\Big]_{\text{SM}}}
\eea
Given the new scalars and VLLs do not modify the $pp\to h$ production rate,
\bea
\mu_{\gamma\gamma} &=&  \frac{\text{BR}(h \to \gamma \gamma)}{\Big[\text{BR}(h \to \gamma \gamma)\Big]_{\text{SM}}}, \\
& \simeq & \frac{\Gamma^{\text{SM}}_{h\to\gamma\gamma}}{\Gamma_{h\to\gamma\gamma}}
\eea 
The latest 13 TeV results on the diphoton signal strength from
the LHC read $\mu_{\gamma\gamma} = 0.99^{+0.14}_{-0.14}$ (ATLAS~\cite{Aaboud:2018xdt}) and
$\mu_{\gamma\gamma} = 1.18^{+0.17}_{-0.14}$ (CMS~\cite{Sirunyan:2018ouh}).
Upon using the standard combination of signal strengths and uncertainties, we obtain $\mu_{\gamma\gamma} = 1.06 \pm 0.1$ and impose this constraint at 2$\sigma$.

\section{Neutrino mass, $\Delta a_{\mu}$ and charged lepton flavour violation} \label{g-2}

We reiterate at the beginning that $\delta^{++}$ and $k^{++}$ respectively couple to only left chiral and right chiral leptons. However, in the mass eigenbasis,
a doubly charged scalar couples to both chiralities. That is, the interactions of muons with the VLLs and 
$H^{++}_{1,2}$ can be expressed as
\bea
\mathcal{L} &=& 2\sum_{i=1,2} \sum_{j=1,2} \bar{\mu}^c
(y_L^{ij} P_L + y_R^{ij} P_R) E_i H_j^{++} + \text{h.c.},
\eea
Where,
\besub
\bea
y_L^{11} &=& y_\Delta^{\mu 4} \text{cos}\a_L \text{cos}\theta, \\
y_R^{11} &=& -y_S^{\mu 4} \text{sin}\a_R \text{sin}\theta, \\
y_L^{12} &=& y_\Delta^{\mu 4} \text{cos}\a_L \text{sin}\theta, \\
y_R^{12} &=& y_S^{\mu 4} \text{sin}\a_R \text{cos}\theta, \\
y_L^{21} &=& y_\Delta^{\mu 4} \text{sin}\a_L \text{cos}\theta, \\
y_R^{21} &=& y_S^{\mu 4} \text{cos}\a_R 
\text{sin}\theta, \\
y_L^{22} &=& y_\Delta^{\mu 4} \text{sin}\a_L \text{sin}\theta, \\
y_R^{22} &=& -y_S^{\mu 4} \text{cos}\a_R 
\text{cos}\theta.
\eea
\eesub

We assume $y_\Delta^{e 4},y_\Delta^{\tau 4},
y_\Delta^{44},y_S^{\a\b},y_S^{e 4},y_S^{\tau 4},
y_S^{44}$ to be vanishingly small\footnote{Demanding $\Delta,k^{++}$ and the VLLs to be odd under some 
$\mathbb{Z}_2$ symmetry while keeping the SM fields even under the same necessitates 
$y_\Delta^{44},y_S^{\a\b},
y_S^{44} = 0.$ We refer to the last paragraph of section \ref{model} for a discussion.}. 
The one-loop muon g-2 $\Delta a_\mu$ has the 
following three distinct components in this limit:
\bea
\Delta a_\mu = \big(\Delta a_\mu^{+}\big)
_{\text{Type-II}}
 + \big(\Delta a_\mu^{++}\big)_{\text{Type-II}}
 + \big(\Delta a_\mu^{++}\big)_{\text{VLL}}.\label{delta_amu}
\eea

In Eq.(\ref{delta_amu}), $\big(\Delta a_\mu^{+}\big)
_{\text{Type-II}}$ ($\big(\Delta a_\mu^{++}\big)
_{\text{Type-II}}$) denotes the contribution from the one-loop amplitude involving SM leptons + singly (doubly) charged scalars. The expression for 
$\big(\Delta a_\mu^{+}\big)
_{\text{Type-II}}$ is given by~\cite{Fukuyama:2009xk}
\besub
\bea
\big(\Delta a_\mu^{+}\big)
_{\text{Type-II}} &=& - \frac{m^2_\mu}{8\pi^2}
\frac{v^2}{v^2 + 2 v^2_\Delta}
\sum_{\a = e,\mu,\tau} \big(y_\Delta^\dagger U_{\text{PMNS}}^* \big)_{\mu \a} 
\big( U^T_{\text{PMNS}} y_\Delta \big)_{\a\mu}\nonumber \\
&&
\int_0^1 dx \frac{x^2(1-x)}{m^2_\mu x^2 + (M^2_{H^+} - m^2_\mu
-m^2_\a)x + m^2_\a} \label{g-2_II_singly}, \\
& \simeq & -\frac{\big(m^2_\nu\big)_{\mu\mu}}{96\pi^2}
\frac{m^2_\mu}{v_\Delta^2 M^2_{H^+}}.
\eea
\eesub
Also,
\besub
\bea
\big(\Delta a_\mu^{++}\big)_{\text{Type-II}} &=&
-\frac{m_\mu^2}{8\pi^2} \sum_{i} \sum_{\a} b_i  
\big( y^\dagger_\Delta \big)_{\mu\a} \big(y_\Delta\big)_{\a\mu} \nonumber \\
&&
\int_0^1 dx \Bigg[ \frac{4x^2(1-x)}{m^2_\mu x^2 + 
((M^{++}_i)^2 - m^2_\mu
-m^2_\a)x + m^2_\a} \nonumber \\
&&
+ \frac{2x^2(1-x)}{m^2_\mu x^2 + (m^2_\a - m^2_\mu
-(M^{++}_i)^2)x + (M^{++}_i)^2} \Bigg], \label{g-2_II_doubly} \\
\text{where} ~b_1 = c^2_\theta,~b_2 = s^2_\theta
\nonumber \\
& \simeq & -\frac{\big(m^2_\nu\big)_{\mu\mu}}{12\pi^2}
\frac{m^2_\mu}{v_\Delta^2} \Bigg( \frac{c^2_\theta}{(M_1^{++})^2} + \frac{s^2_\theta}{(M_2^{++})^2} \Bigg).
\eea
\eesub
The contribution from to $\Delta a_\mu$ from $H^+$ is identical to the minimal Type-II seesaw.
In fact, the contribution from doubly charged scalars is also qualitatively the same as can be checked from~\cite{Fukuyama:2009xk}. In either case, the scalars only couple to the left-chiral components of the SM fermions and hence no chirality flip occurs in the muon g-2 amplitudes. Most importantly, one finds that 
$\big(\Delta a_\mu^{++}\big)_{\text{Type-II}} < 0$ implying that the Type-II-like amplitudes cannot explain the muon anomaly~\cite{Fukuyama:2009xk}. 

The contribution from the VLLs is\footnote{An excellent review containing analytical formulae for $\Delta a_\mu$  for different classes of models is \cite{Lindner:2016bgg}}
\bea
\big(\Delta a_\mu^{++}\big)_{\text{VLL}} &=&
\sum_{i=1,2} \sum_{j=1,2} \Bigg[ -\frac{m^2_\mu}{4\pi^2} \Big(\{(y_L^{ij})^2 + (y_L^{ij})^2\} I_1(M_i,M_j^{++}) + \frac{2 M_i}{m_\mu} y_L^{ij} y_R^{ij} I_2(M_i,M_j^{++}) \Big) \nonumber \\
&&
-\frac{m^2_\mu}{2\pi^2} \Big(\{(y_L^{ij})^2 + (y_R^{ij})^2\} I_3(M_i,M_j^{++}) + \frac{2 M_i}{m_\mu} y_L^{ij} y_R^{ij}
I_4(M_i,M_j^{++}) \Big) \Bigg].
\label{g-2_VLL}
\eea
The integrals $I_{a}(m_1,m_2),~a=1,2,3,4$ upon neglicting $m_\mu$ are
\besub
\bea
I_1(m_1,m_2) &=& \int_0^1 dx~\frac{x^2(1-x)}{m_1^2 x + m_2^2(1-x)}, \\
I_2(m_1,m_2) &=& \int_0^1 dx~\frac{x^2}{m_1^2 x + m_2^2(1-x)}, \\
I_3(m_1,m_2) &=& \int_0^1 dx~\frac{x^2(1-x)}{m_1^2 (1-x) + m_2^2 x}, \\
I_4(m_1,m_2) &=& \int_0^1 dx~\frac{x(1-x)}{m_1^2 (1-x) + m_2^2 x}.
\eea
\eesub
The integrals $I_{a}(m_1,m_2),~a=1,2,3,4$ are all positive and their analytical expressions are given in the Appendix. It then follows that the contribution to $\big(\Delta a_\mu^{++}\big)_{\text{VLL}}$
from the first and third terms in Eq.(\ref{g-2_VLL}) are negative.  
In contrast, a chirality flip is noted in the second and fourth terms. To examine this contribution more closely, we define $\Delta M \equiv M_2 - M_1 << M_1$ and $\Delta M^{++} \equiv M_2^{++} - M_1^{++} << M_1^{++}$, and, take $\a_L = \a_R$ for simplicity. The chirality-flipping contribution in its lowest order of $\Delta M$ and $\Delta M^{++}$ then becomes
\besub
\bea
\big(\Delta a_\mu^{++}\big)_{\text{VLL}}^{\text{cf}} &\simeq& \frac{m_\mu}{4\pi^2}y_{\Delta}^{\mu 4}y_{S}^{\mu 4} \text{sin}(2\a_R)\text{sin}(2\theta) \frac{\Delta M \Delta M^{++}}{(M_1^{++})^3} f\Big( \frac{M_1^2}{(M_1^{++})^2} \Big), \label{cf} \\
f(r) &=& \frac{(-35 r^3 + 15 r^2 + 27 r - 7) + (12 r^3 + 40 r^2 - 2 r - 1)\text{log}(r)}{2(r-1)^5}.
\eea
\eesub
Thus, non-zero mass splittings between $E_1,E_2$ and $H_1^{++},H_2^{++}$ and correspondingly non-zero mixings are necessary to achieve a non-zero chirality-flip for 
$\a_L = \a_R$\footnote{This observation highlights the role of EWSB and the parameters $\l_8,y_4,y_4^\prime$ in generating the mass splittings and ultimately predicting the observed value of $\Delta a_\mu$.}. More importantly, Eq.(\ref{cf}) shows that the chirality-flipped amplitude can be of either sign. In fact, it is enhanced \emph{w.r.t.} the chirality preserving part of $\big(\Delta a_\mu^{++}\big)_{\text{VLL}}$ and the Type-II like terms by an 
$\mathcal{O} \Big(\frac{M_i}{m_\mu} \Big)$ factor. It is therefore possible to generate a positive contribution of the requisite magnitude by choosing the parameters appropriately.  

\begin{figure}[tbhp]
\begin{center}
\includegraphics[scale=0.50]{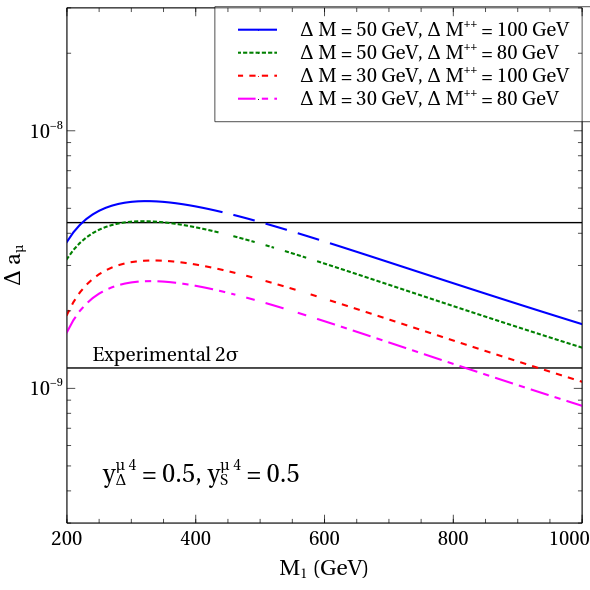}~~~~
\includegraphics[scale=0.50]{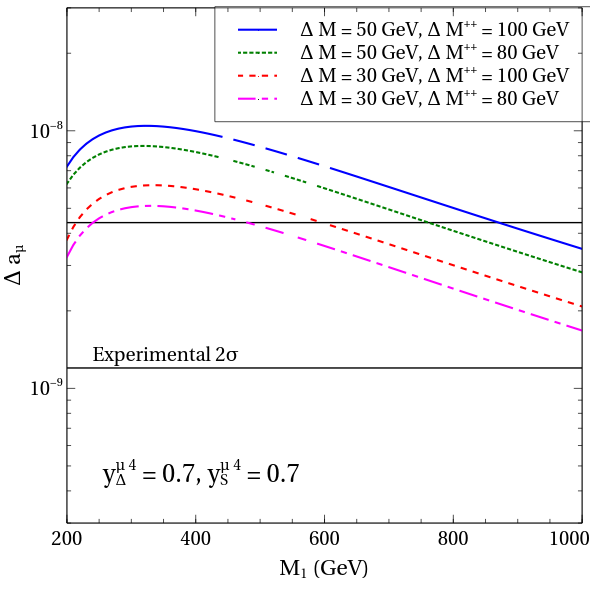}
\caption{The variation of $\Delta a_\mu$ with $M_1$ for different values of $\Delta M$ and $\Delta M^{++}$ and two sets of 
$\big(y_\Delta^{\mu 4},y_S^{\mu 4}\big)$. The horizontal straight lines correspond to the observed 2$\sigma$ limit. The values chosen for the other parameters are given in the text. The color-coding is explained in the legends.}
\label{f:g-2}
\end{center}
\end{figure}

To numerically test the chirality-flipping effect, we plot $\Delta a_\mu$ versus $M_1$
in FIG.~\ref{f:g-2} for $\Delta M = M_2 - M_1$ = 30 GeV, 50 GeV; $\Delta M^{++} = M^{++}_2 - M^{++}_1$ = 80 GeV, 100 GeV; and; 
$\big(y_\Delta^{\mu 4},y_S^{\mu 4}\big)$ = (0.5,0.5),
 (0.7,0.7). The values chosen for the other parameters are $v_\Delta = 10^{-3} ~\text{GeV}, M_\Delta = M_S = M_{H^+} = M_1^{++} = 500~\text{GeV},
\theta = \a_L = \a_R = \frac{\pi}{4}$. 

An inspection of FIG.~\ref{f:g-2} ascertains that the aforementioned chirality-flip can indeed lead to an explanation of the muon anomaly in  this model. We reiterate that in $\big(\Delta a_\mu^{++}\big)_{\text{VLL}}$, the chirality-flipped contribution 
is enhanced \emph{w.r.t} the negative terms by 
$\mathcal{O}(M_i/m_\mu)$. Therefore, FIG.~\ref{f:g-2} essentially captures the behaviour of the chirality-flipped amplitude. It is seen that the larger are the mass splittings $\Delta M$ and $\Delta M^{++}$, the larger is the size of chirality-flip and hence, the larger is $\Delta a_\mu$. Though Eq.(\ref{cf}) is derived for 
$\Delta M << M_1,~\Delta M^{++} << M_1^{++}$, it still intuitively indicates a larger muon g-2 value for larger mass splittings, thereby explaining the said behaviour in FIG.~\ref{f:g-2}. Eq.(\ref{cf}) also shows that the chirality-flip is proportional to the product 
$y_\Delta^{\mu 4} y_S^{\mu 4}$ and this explains the higher in $\Delta a_\mu$ in the right plot compared to
the left for a given set of $\Delta M$, $\Delta M^{++}$ and the mixing angles.

\begin{figure}[tbhp]
\begin{center}
\includegraphics[scale=0.50]{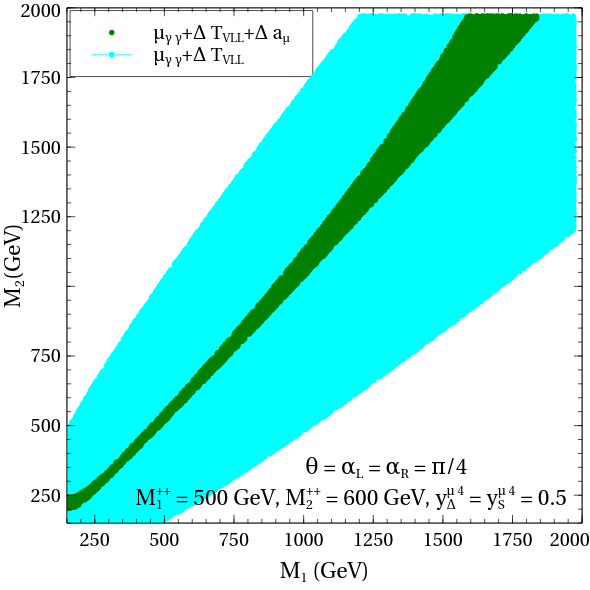}~~~~
\includegraphics[scale=0.50]{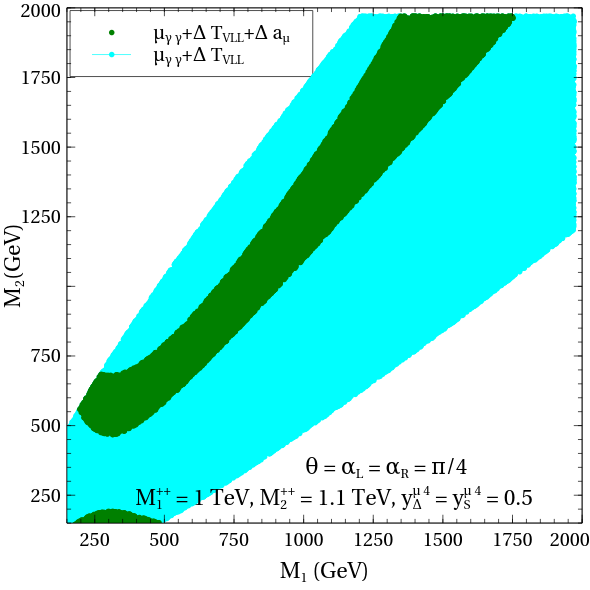} \\
\includegraphics[scale=0.50]{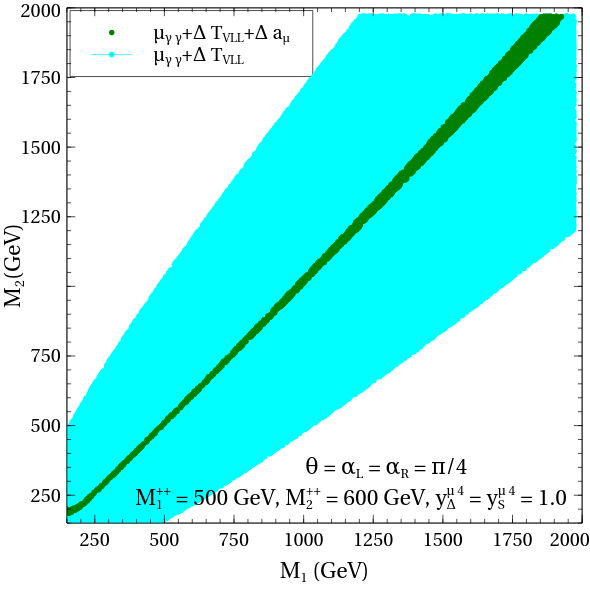}~~~~
\includegraphics[scale=0.50]{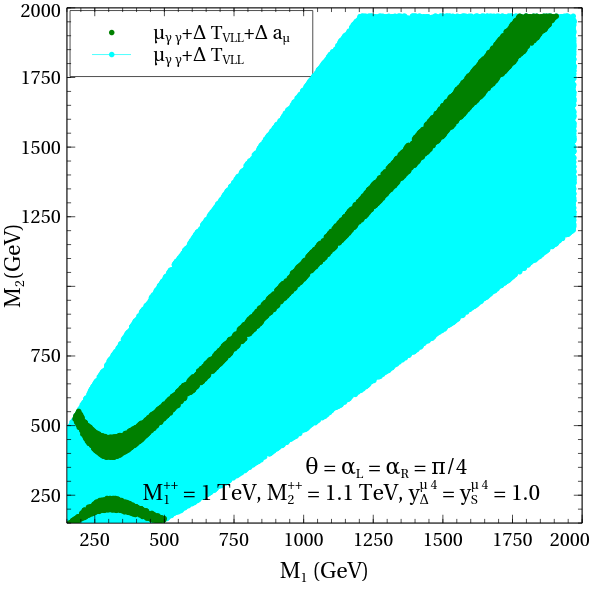} \\
\caption{Parameter region in the $M_1-M_2$ plane allowed by the diphoton and $T$-paramater constraint (sky blue), and,
leading to $\Delta a_\mu$ in the observed 2$\sigma$ limit in addition to
satisfying the diphoton and $T$-parameter constraints (green).
The values taken by the other parameters are shown in the plots.}
\label{f:M1_M2}
\end{center}
\end{figure}

The chirality flip is further probed by identifying the region in the $M_1-M_2$ plane leading to the observed $\Delta a_\mu$. FIG.~\ref{f:M1_M2} shows the parameter region allowed by the diphoton and $T$-parameter constraints for specific choices for the other relevant parameters (as seen in the plots). A smaller region is seen to explain the muon anomaly for each case.
FIG.~\ref{f:M1_M2} too can be interpreted using Eq.(\ref{cf}). With $\Delta M^{++}$ = 100 GeV for each panel,
as $M_1^{++}$ increases from 500 GeV to 
1 TeV keeping the Yukawa couplings and the mixing angles fixed, the denominator of $\big(\Delta a_\mu^{++}\big)_{\text{VLL}}^{\text{cf}}$ increases and hence
$\Delta M$ must accordingly increase to maintain $\Delta a_\mu$ in the 2$\sigma$ band. This is precisely why the band expands upon switching from the top left to the top right panel. For example, with $M_2$ = 2 TeV, 
$M_1 \in $ [1.59 TeV,1.84 TeV] expands to $M_1 \in $ [1.35 TeV,1.75 TeV] here. The shrinkage seen while switching from top left to bottom left, i.e., from $y_{\Delta}^{\mu 4} = y_S^{\mu 4}$ = 0.5 to 1, is also expected since increasing 
$y_{\Delta}^{\mu 4} y_S^{\mu 4}$ while keeping the other parameters fixed would cause $\Delta M$ to appropriately constrict ($M_1 \in $ [1.85 TeV,1.93 TeV], correspondingly).
\begin{figure}[tbhp]
\begin{center}
\includegraphics[scale=0.50]{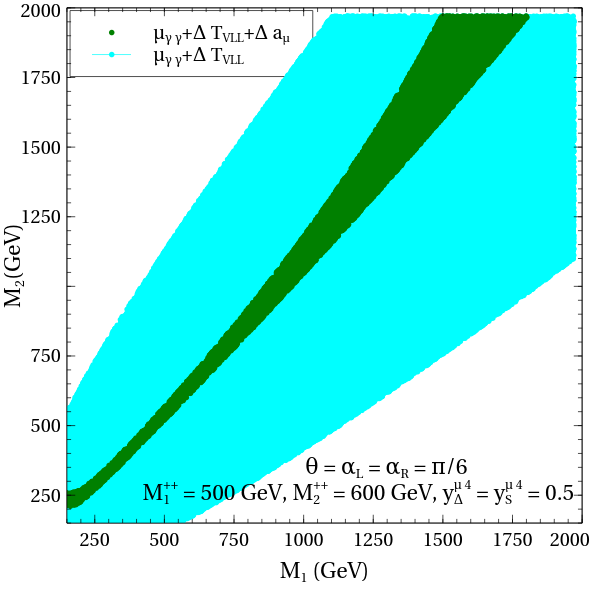}~~~~
\includegraphics[scale=0.50]{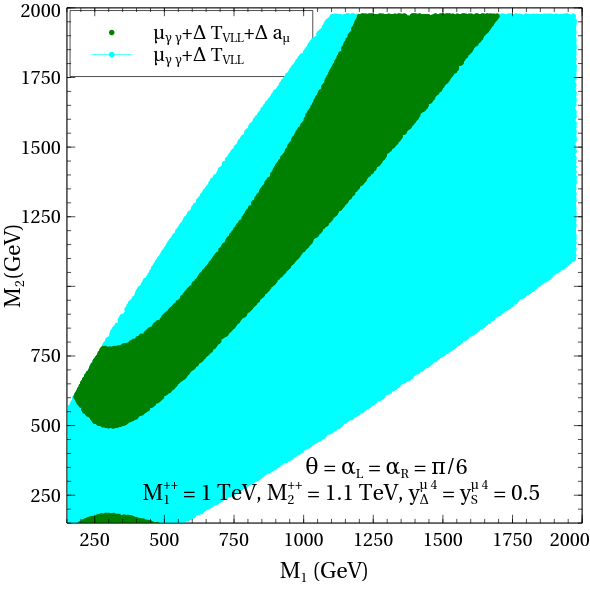} \\
\includegraphics[scale=0.50]{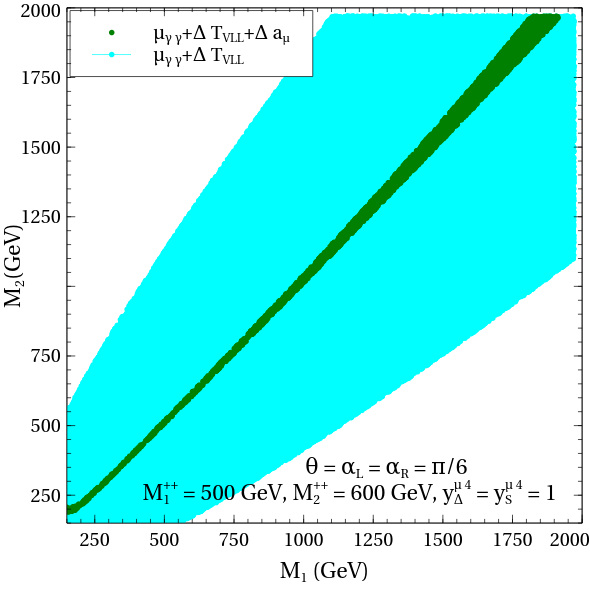}~~~~
\includegraphics[scale=0.50]{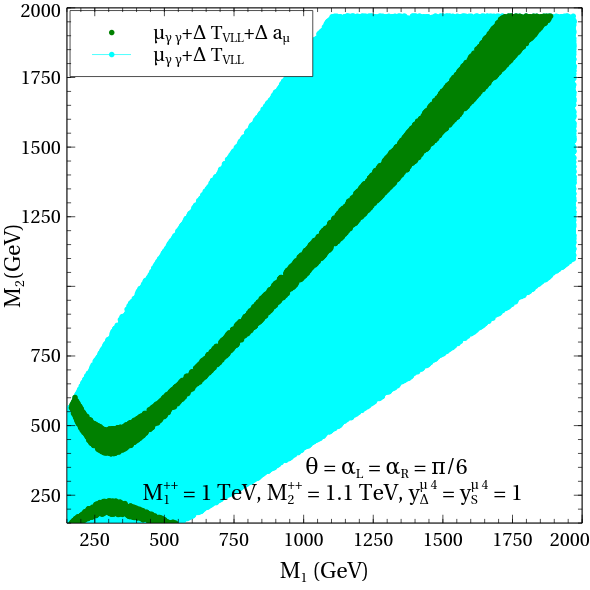} \\
\caption{Same as FIG.~\ref{f:M1_M2} but with $\theta = \a_L = \a_R = \frac{\pi}{6}$. }
\label{f:M1_M2_piby6}
\end{center}
\end{figure}
One also reads from Eq.(\ref{cf}) that $\theta=\a_L=\a_R = \frac{\pi}{4}$ maximises $\big(\Delta a_\mu^{++}\big)_{\text{VLL}}^{\text{cf}}$ for mixed massses and Yukawa couplings. Changing the value to $\frac{\pi}{6}$ therefore implies a more relaxed band in the $M_1-M_2$ plane compared to the corresponding one for $\frac{\pi}{4}$. This is concurred by an inspection of FIG.~\ref{f:M1_M2_piby6}. Each band in this case is broader than the corresponding one for $\frac{\pi}{4}$.

The $\mathcal{O}(M_i/m_\mu)$ chirality-flipped enhancement does not occur in the $l_\a \to l_\b \gamma$ amplitudes~\cite{Lavoura:2003xp} for $\a \neq \b$ due to the assumption that 
$y_\Delta^{e 4},y_\Delta^{\tau 4}$, 
$y_S^{\b 4}$ are vanishingly small\footnote{Even if no such approximation is a priori made, an estimation of the LFV chirality-flipping amplitude using \cite{Lindner:2016bgg} leads to $|y_{\Delta}^{\a 4}|, |y_S^{\a 4}| \sim \mathcal{O}(10^{-4})$ for $\a = e,\tau$ here.}. Analogously to $\big(\Delta a_\mu^{+}\big)_{\text{Type-II}}$ and $\big(\Delta a_\mu^{++}\big)_{\text{Type-II}}$,
a non-zero 
$l_a \to l_\b \gamma$ amplitude
is therefore induced only by the triplet $\Delta$ that couples to only the left-chiral components of the SM leptons. The $l_\a \to l_\b \gamma$ amplitude is then qualitatively similar to that in the minimal Type-II case~\cite{Akeroyd:2009nu}. One then finds the corresponding branching ratio in the present model to be 
\bea
\text{BR}(l_\a \to l_\b \gamma) &=& \frac{\a_{em} 
|(m_\nu^2)_{\a\b}|^2}
{12\pi G_F^2 v_\Delta^4} \Bigg(\frac{1}{8 M^2_{H^+}}
 + \frac{c^2_\theta}{(M_1^{++})^2}
 + \frac{s^2_\theta}{(M_2^{++})^2} \Bigg)^2 \text{BR}(l_\a \to l_\b \nu_\a \bar{\nu_\b})\label{ltolgamma}.
\eea

Similarly, the branching ratios of the 3-body CLFV decays are given by \footnote{The corresponding formula for the Higgs triplet model is seen in \cite{Akeroyd:2009nu,Akeroyd:2009hb}}
\besub
\bea
\text{BR}(\mu \to \bar{e} e e) &=& \frac{|m_\nu^{e\mu}|^2 |m_\nu^{ee}|^2}{16 G^2_F v^4_\Delta} \Bigg(\frac{c^2_\theta}{(M_1^{++})^2}
 + \frac{s^2_\theta}{(M_2^{++})^2} \Bigg)^2 
\text{BR}(\mu \to e \bar{\nu_e} \nu_\mu), \\
\text{BR}(\tau \to \bar{l_{\a}} l_{\b} l_{\gamma}) &=& S\frac{|m_\nu^{\tau \a}|^2 |m_\nu^{\b \gamma}|^2}{16 G^2_F v^4_\Delta} \Bigg(\frac{c^2_\theta}{(M_1^{++})^2}
 + \frac{s^2_\theta}{(M_2^{++})^2} \Bigg)^2 
\text{BR}(\tau \to \mu \bar{\nu_\mu} \nu_\tau).
\eea
\eesub
In the above, $S$ = 1(2) for $\beta=\gamma$ ($\beta \neq \gamma$),
$\text{BR}(\mu \to e \bar{\nu_e} \nu_\mu)$ = 100$\%$ and $\text{BR}(\tau \to \mu \bar{\nu_\mu} \nu_\tau)$ = 17$\%$. The updated CLFV bounds are summarised in Table~\ref{lfv_bound}.

\begin{table}
\centering
\begin{tabular}{ |c|c| } 
\hline
 LFV channel & Experimental bound \\ 
 \hline \hline 
  $\mu \rightarrow e \gamma$ & $<$ 4.2 $\times 10^{-13}$ 
 ~\cite{TheMEG:2016wtm} \\ \hline
   $\tau \rightarrow e \gamma$ & $<$ 1.5 $\times 10^{-8}$ 
 ~\cite{Aubert:2009ag} \\ \hline
   $\tau \rightarrow \mu \gamma$ & $<$ 1.5 $\times 10^{-8}$ 
 ~\cite{Aubert:2009ag} \\ \hline
  $\mu \rightarrow \bar{e} e e$ & $<$ 1 $\times 10^{-12}$ 
 ~\cite{Bellgardt:1987du} \\ \hline
  $\tau \rightarrow \bar{e} e e$ & $<$ 1.4 $\times 10^{-8}$ 
 ~\cite{Amhis:2016xyh} \\ \hline
 $\tau \rightarrow \bar{\mu} e e$ & $<$ 8.4 $\times 10^{-9}$ 
 ~\cite{Amhis:2016xyh} \\ \hline
  $\tau \rightarrow \bar{\mu} e \mu$ & $<$ 1.6 $\times 10^{-8}$ 
 ~\cite{Amhis:2016xyh} \\ \hline
  $\tau \rightarrow \bar{e} \mu \mu$ & $<$ 9.8 $\times 10^{-9}$ 
 ~\cite{Amhis:2016xyh} \\ \hline
   $\tau \rightarrow \bar{e} \mu e$ & $<$ 1.1 $\times 10^{-8}$ 
 ~\cite{Amhis:2016xyh} \\ \hline
   $\tau \rightarrow \bar{\mu} \mu \mu$ & $<$ 1.2 $\times 10^{-8}$ 
 ~\cite{Amhis:2016xyh} \\ \hline
\end{tabular}
\caption{Latest upper limits on LFV branching ratios.}
\label{lfv_bound}
\end{table}
We read from Eq.(\ref{ltolgamma}) that the size of such branching ratios for all 
$\a,\b=e,\mu,\tau$ is controlled by $v_\Delta$ for fixed scalar masses. Choosing an appropriately large $v_\Delta$ therefore suffices to evade the CLFV bounds.

\section{Analysis combining electroweak vacuum stability} \label{vacstab}
In this section, we look for a stable EW vacuum till the Planck scale within the parameter space compatible with the observed muon g-2. The boundary scale, or the scale from which the couplings begin to evolve towards high scales 
is chosen to be the $t$-pole mass, i.e., $M_t$ = 173.34 GeV. We first note the following additional terms in the $\beta$-function of the Higgs quartic coupling $\l$ \emph{w.r.t.} the SM.  
\bea
\beta_\l = \beta_\l^{\text{SM}} + 6\l_4^2 + 4\l_5^2 + 2\l_6^2 + 4 \l_8^2 + 4\l \big(y_4^2 + (y_4^\prime)^2 \big)
- 4 y_4^4 - 4 (y_4^\prime)^4
\eea
A complete set of the one-loop beta functions is given in the Appendix. Those for the Yukawa couplings $y_\Delta^{e 4},y_\Delta^{\tau 4},
y_\Delta^{44},y_S^{\a\b},y_S^{e 4},y_S^{\tau 4}$ and
$y_S^{44}$ are however neglected since, for instance, $y_\Delta^{e 4} \to 0$ at the EW scale implies $y_\Delta^{e 4} \to 0$ at all scales. The presence of both bosonic and fermionic terms in $\beta_\l - \beta_\l^{\text{SM}}$ paves the path for an interesting interplay. We throughout take $M_\Delta = M_S = M_H^+ = M_1^{++}$ as well as $\l_1 = \l_2 = \l_7 = 0.01, \l_3 = 0.3$ at the boundary scale for simplification. We also take $M_W$ = 80.384 and $\a_s(M_Z)$ = 0.1184 in which case the $t$-Yukawa and the gauge couplings at the boundary scale are 
$y_t(\mu=M_t) = 0.93690,~g_1(\mu=M_t) = 0.3583,
~g_2(\mu=M_t) = 0.6478,
~g_3(\mu=M_t) = 1.1666$~\cite{Buttazzo:2013uya}.

\begin{table}
\centering
\begin{tabular}{ |c|c|c| } 
\hline
 & BP1 & BP2 \\ 
\hline
$v_\Delta$ & $10^{-3}$ GeV & $10^{-8}$ GeV\\
$M_1$ & 850.0 GeV & 200.0 GeV\\
$M_2$ & 920.8 GeV & 236.8 GeV\\
$M_1^{++}$ & 200.0 GeV & 800.0 GeV\\
$M_2^{++}$ & 270.0 GeV & 854.4 GeV\\
$y_\Delta^{\mu 4}$ & 0.497 & 0.447\\
$y_S^{\mu 4}$ & 0.345 & 0.430\\
$\theta$ & 0.158 & 0.100\\
$\alpha_L$ & 0.732 & -1.209\\
$\alpha_R$ & 0.760 & -0.955 \\ \hline
$\Delta a_\mu$ & $1.372 \times 10^{-9}$ & $1.601 \times 10^{-9}$\\
BR$(\mu \to e \gamma)$ & $8.562 \times 10^{-35}$ & $3.404 \times 10^{-17}$ \\
BR$(\tau \to e \gamma)$ & $8.562 \times 10^{-35}$ & $6.064 \times 10^{-18}$ \\
BR$(\tau \to \mu \gamma)$ & $8.562 \times 10^{-35}$ & $1.065 \times 10^{-16}$ \\
BR$(\mu \to \bar{e} e e)$ & $2.657 \times 10^{-35}$ & $1.059 \times 10^{-17}$ \\
BR$(\tau \to \bar{e} e e)$ & $6.733 \times 10^{-36}$ & $2.683 \times 10^{-18}$ \\
BR$(\tau \to \bar{\mu} e e)$ & $7.224 \times 10^{-35}$ & $2.879 \times 10^{-17}$ \\
BR$(\tau \to \bar{e} \mu e)$ & $1.197 \times 10^{-34}$ & $4.774 \times 10^{-17}$ \\
BR$(\tau \to \bar{e} \mu \mu)$ & $1.636 \times 10^{-33}$ & $6.521 \times 10^{-16}$ \\
BR$(\tau \to \bar{\mu} e \mu)$ & $1.285 \times 10^{-33}$ & $5.122 \times 10^{-16}$ \\
BR$(\tau \to \bar{\mu} \mu \mu)$ & $1.755 \times 10^{-32}$ & $6.997 \times 10^{-15}$ \\
$\Delta T_{\text{VLL}}$ & $\simeq$ 0.006 & $\simeq$ 0.003 \\
$\mu_{\gamma\gamma}$ & 0.983 & 0.930 \\ \hline
\end{tabular}
\caption{Benchmarks to demonstrate EW vacuum stability in the present scenario.}
\label{BP}
\end{table}
We propose two benchmarks in Table~\ref{BP} in order to gain insight on the evolution under RG. These benchmarks
pass all the relevant constraints and predict
$\Delta a_\mu$ in the 2$\sigma$ range.  
\begin{figure}[tbhp]
\begin{center}
\includegraphics[scale=0.57]{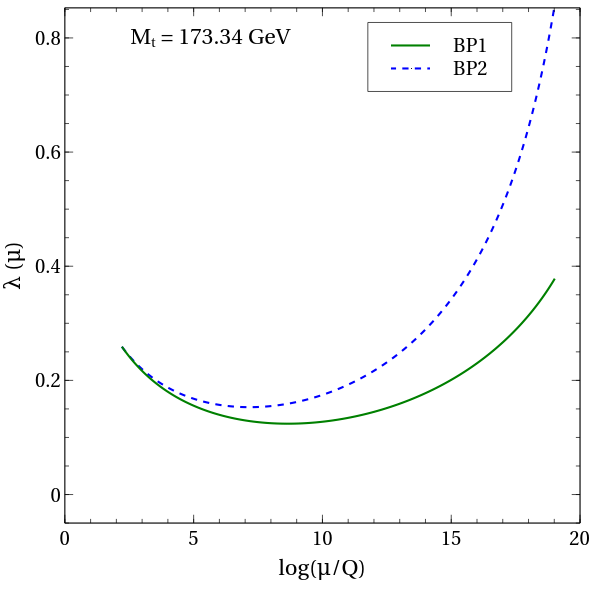} \\ 
\includegraphics[scale=0.57]{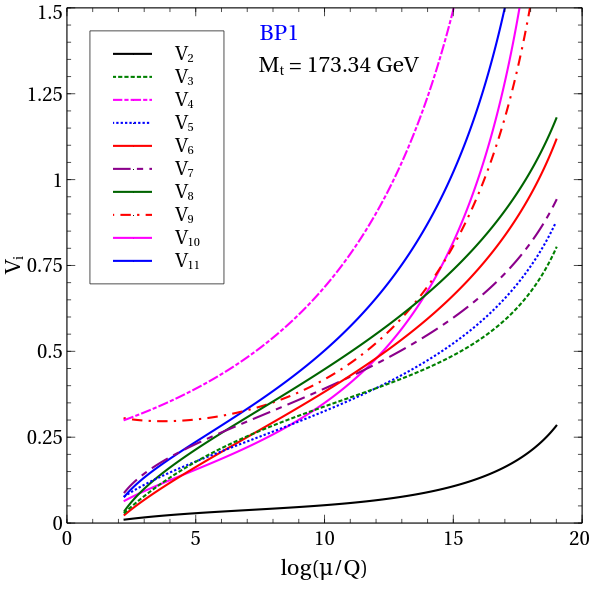}~~~
\includegraphics[scale=0.57]{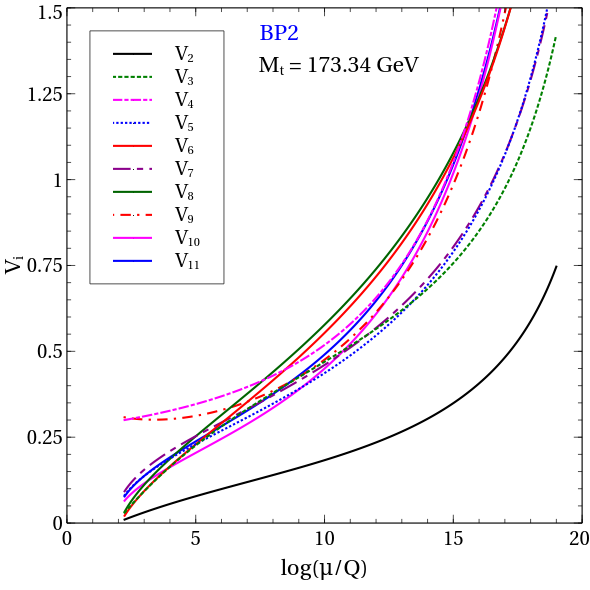}~~~
\caption{Top figure: RG evolution of $\l$ for BP1 and BP2. Bottom left (right) figure: RG evolutions of $V_i$ for BP1 (BP2). The color-coding is explained in the legends.}
\label{f:BP}
\end{center}
\end{figure} 
FIG.~\ref{f:BP} displays the RG running of $\l$ and 
$V_i$
for BP1 and BP2. Both the benchmarks are seen to offer a stable EW vacuum till the Planck scale (taken to be $\simeq 10^{19}$ GeV) since $\l$ is rendered positive throughout the evolution. 
Besides, one also finds $V_i > 0$ for either benchmark.
It is worthwhile to comment on the role of $\l_8$ in stabilising the vacuum. For both BP1 and BP2, $\l_4(M_t)$ = 0 and $\l_5(M_t),\l_6(M_t) = \mathcal{O}(0.01)$. Such small values do not suffice to ensure $\l > 0$ till the Planck scale and considering that $\l_{4,5,6}$ have gentle RG evolution trajectories, it is actually $\l_8$ that stabilises the EW vacuum. That $\l$ in case of BP2 increases more rapidly under RG compared to BP1 is also attributed to the different $\l_8(M_t)$ values in the two cases. While $\l_8(M_t)$ = 0.169 for BP1, it equals 0.297
for BP2 implying a stronger bosonic push to the RG evolution of $\l$ in case of the latter. And, for either benchmark, the fermionic contribution coming from $y_4$ and $y_4^\prime$ is too weak to counter the bosonic effect. Therefore, it is established that with an appropriate choice of the parameters, the explanation of the muon anomaly in the current scenario complies with a stable EW vacuum till the Planck scale. 

\begin{figure}[tbhp]
\begin{center}
\includegraphics[scale=0.50]{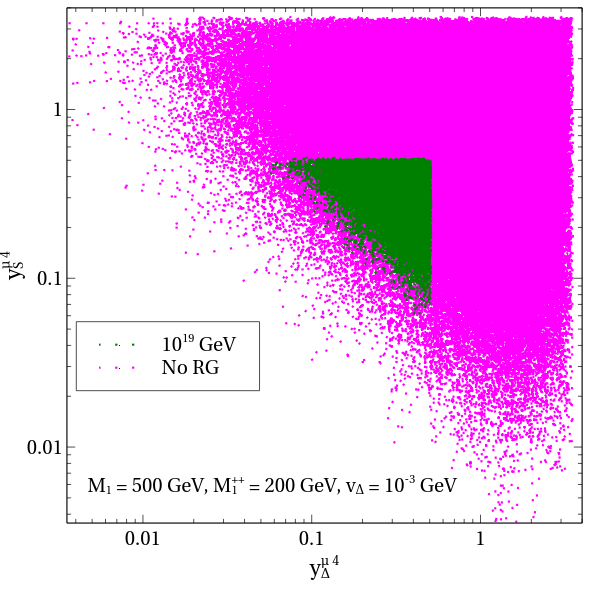}~~~ 
\includegraphics[scale=0.50]{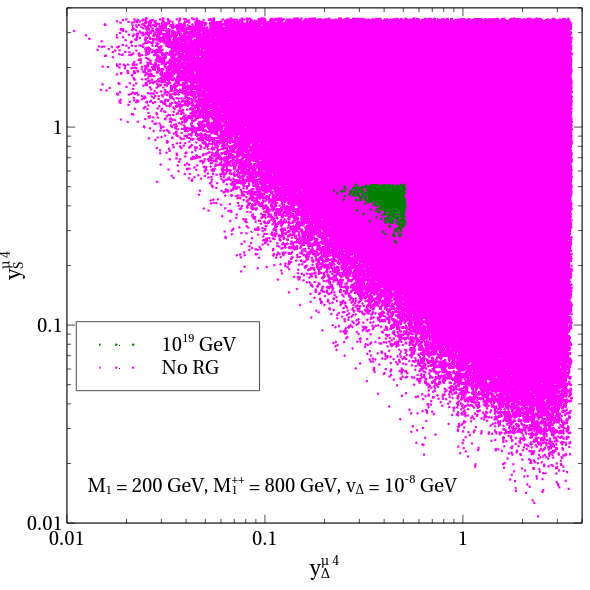}
\caption{Allowed parameter points in the 
$y_\Delta^{\mu 4}$-$y_S^{\mu 4}$ plane. The 
magenta points satisfy all requirements except vacuum stability while the green points additionally ensure a stable vacuum and a perturbative theory till $10^{19}$ GeV.}
\label{f:ydel_yS}
\end{center}
\end{figure}

\begin{figure}[tbhp]
\begin{center}
\includegraphics[scale=0.50]{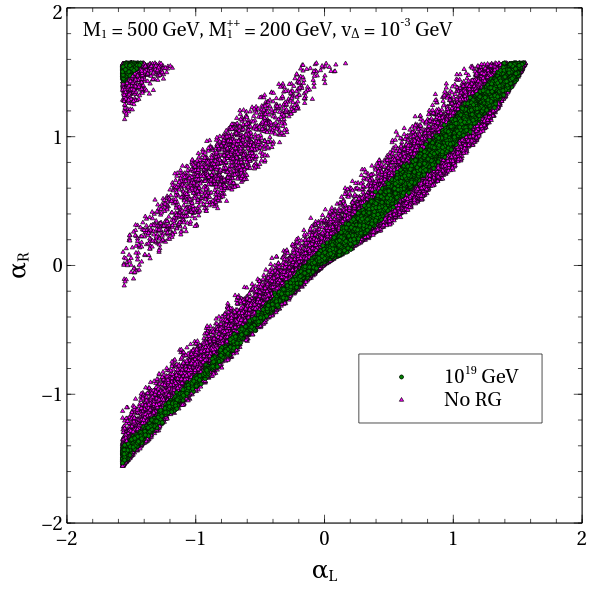}~~~ 
\includegraphics[scale=0.50]{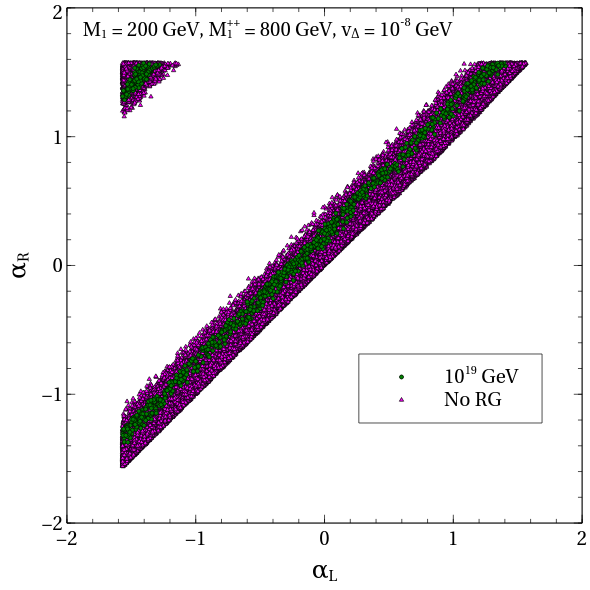}
\caption{Allowed parameter points in the 
$\a_L$-$\a_R$ plane. The color-coding is the same as in 
FIG.~\ref{f:ydel_yS}.}
\label{f:aL_aR}
\end{center}
\end{figure}

\begin{figure}[tbhp]
\begin{center}
\includegraphics[scale=0.50]{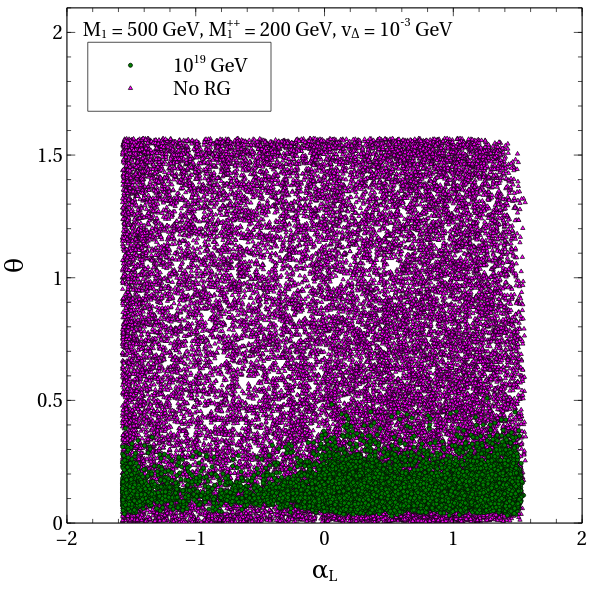}~~~ 
\includegraphics[scale=0.50]{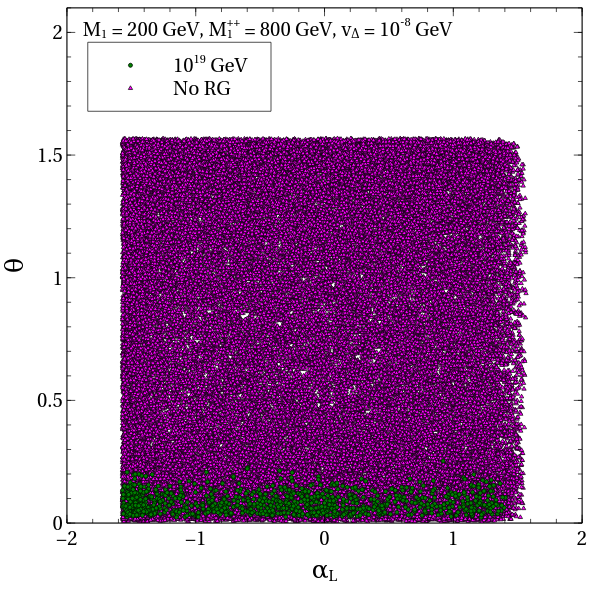}
\caption{Allowed parameter points in the 
$\a_L$-$\theta$ plane. The color-coding is the same as in 
FIG.~\ref{f:ydel_yS}.}
\label{f:aL_t}
\end{center}
\end{figure}

Next, we try to extract parameter regions consistent with all the constraints, a value of $\Delta a_\mu$ within the 2$\sigma$ range as well as a stable EW vacuum till the Planck scale. We choose the set ($v_\Delta,M_1,M_1^{++}$) to be ($10^{-3}$ GeV, 500 GeV, 200 GeV) and ($10^{-8}$ GeV, 200 GeV, 800 GeV) make the following variation of the rest  parameters: 
\besub
\bea
0 \leq \Delta M \leq 100~\text{GeV},~~0 \leq \Delta M^{++} \leq 100~\text{GeV}, \\ 
0 \leq y_\Delta^{\mu 4}, y_S^{\mu 4} \leq \sqrt{4\pi}, \\
0 \leq \theta \leq \frac{\pi}{2},~~-\frac{\pi}{2} \leq \a_L,\a_R \leq \frac{\pi}{2}. 
\eea
\eesub
A parameter point is selected if it clears all the constraints and leads to a muon g-2 value within 2$\sigma$. Further, all such parameter points are evolved under RG and a subset yielding a stable vacuum and also abiding by perturbativity and unitarity up to $\mu = 10^{19}$ GeV is identified. The parameter points are plotted in the 
$y_\Delta^{\mu 4}$-$y_S^{\mu 4}$ (FIG.~\ref{f:ydel_yS}), 
$\a_L$-$\a_R$ (FIG.~\ref{f:aL_aR}) and $\a_L$-$\theta$
(FIG.~\ref{f:aL_t}) planes.

We find upon inspecting FIG.~\ref{f:ydel_yS} that the requirement of validity till high scales greatly constrains $y_\Delta^{\mu 4}$ and $y_S^{\mu 4}$. In fact, |$y_\Delta^{\mu 4}$|, |$y_\Delta^{\mu 4}$| 
$\lesssim$ 0.5. Above this value, these Yukawa couplings become non-perturbative at scales lower than the Planck scale irrespective of the values taken by the other parameters. Therefore, an upper bound is derived from
high scale perturbativity. However, the lower bound depends on the choice of the other parameters. For instance, the lower bound $y_\Delta^{\mu 4} \gtrsim 0.21,~y_S^{\mu 4} \gtrsim 0.25$ for the 
$v_\Delta = 10^{-3},M_1=500~\text{GeV},M_1^{++}=200~\text{GeV}$ configuration is more stringent than that obtained for
$v_\Delta = 10^{-8}~\text{GeV},M_1=200~\text{GeV},M_1^{++}=800~\text{GeV}$, i.e. $y_{\Delta}^{\mu 4}$, $y_S^{\mu 4} \gtrsim 0.06$.

FIG.~\ref{f:aL_aR} shows that in the $\a_L>0,\a_R>0$ quadrant, the magenta region essentially comprises a band about the $\a_R = \a_L$ straight line.
Demanding validity till the Planck scale constricts the band further. This can be traced to the fact that validity till high scales favours $y_4 \sim y_4^\prime$. And for the $(M_1,M_2)$ values
taken in the scan, $y_4 \sim y_4^\prime$ favours $\a_L \sim \a_R$
(see Eqs.(\ref{y4}) and (\ref{y4p})). In addition, as corroborated by FIG.~\ref{f:aL_t}, a stable vacuum and a pertubative theory till $10^{19}$ GeV imposes an upper bound on $\theta$. This is expected
since a constraint on $\l_8$ from vacuum stability and perturbative unitraity shall always translate to a corresponding constraint on 
$\theta$ (see Eq.(\ref{l8})). The bound, i.e., $\theta \leq 0.25$ is more restrictive for $v_\Delta = 10^{-8}~\text{GeV},M_1=200~\text{GeV},M_1^{++}=800~\text{GeV}$ than the corresponding $\theta <0.50$ for $v_\Delta = 10^{-3}~\text{GeV},M_1=500~\text{GeV},M_1^{++}=200~\text{GeV}$.

\section{Summary and conclusions}\label{summary}

If the Run 1 data of the "MUON G-2" experiment \cite{fnal:2021gmt} corroborates
the existing discrepancy, the hint of NP contributing to the muon anomalous magnetic moment will get stronger. The present study puts forth one such NP scenario. In this work, we have extended the minimal Type-II seesaw scenario by a doubly charged scalar singlet, an $SU(2)_L$ doublet of vector-like leptons and, a charged $SU(2)_L$ singlet vector-like lepton. While mixing between the newly introduced vector leptons and the SM leptons is neglected, it is allowed between the VLLs themselves. Similarly, the scalar potential allows for a mixing between the two doubly charged scalars in the framework. Therefore, the doubly charged mass eigenstates  couple to both chiralities of leptons. We have explained how a chirality-flip can then explain the observed value of the muon g-2, something not possible within the minimal Type-II model alone. A non-zero neutrino mass and appropriately suppressed CLFV can be predicted at the same time. Another pertinent issue the present work touches upon is that of EW vacuum stability given that the scenario features additional scalar degrees of freedom. We have computed the one-loop RG equations for this model and demonstrated that the parameter region accounting for a $\Delta a_\mu$ value in the desired range can also lead to a stable EW vacuum up to the Planck scale. An interesting follow-up constitutes engineering a similar chirality-flip for $\Delta a_\mu$ connecting the doubly charged scalars coming from the LRSM and such an investigation is presently underway \cite{Chakrabarty:2022ktnc}. 

The following lepton-rich signals can arise at the LHC for this model whenever $M_1 > M^{++}_{1,2}$:
\begin{itemize}

\item $pp \to E_1^+ E_1^- \to H_i^{\pm\pm} H_j^{\mp\mp} l^+ l^- \to 6l$,

\item $pp \to E_1^+ E_1^- \to H_i^{\pm\pm} H_j^{\mp\mp} l^+ l^- \to W^+ W^- W^+ W^- l^+ l^- \to 6l + \met$ for $i,~j$ = 1,2.

\end{itemize}

We assumed that $H_i^{\pm\pm}$ dominantly decays to $l^\pm l^\pm$ and $W^\pm W^\pm$ for the first and second cascade respectively. For either case, demanding a total of 6 leptons in the final state can definitely help suppress the SM background. Moreover, the first signal is not accompanied by neutrinos and hence the doubly charged scalar masses are fully reconstructible modulo the combinatorics. A successful reconstruction of the scalar masses in the $i \neq j$ case confirms the presence of two distinct doubly charged scalars thereby distinguishing this scenario from the minimal Type-II model, at colliders.

\acknowledgments
NC acknowledges financial support from Indian Institute of Science through the C V Raman Post-Doctoral fellowship. He also acknowledges support from DST, India, under grant number IFA19-PH237 (INSPIRE Faculty Award).

\section{Appendix}

\subsection{Unitarity}
We compute here the $2\to2$ scattering matrices and their eigenvalues in the basis of two-particle states. 

\textbf{Neutral 2-particle states:} 
We take the basis as 
$\{\delta^+ \phi^-, \phi^+ \delta^-, \delta^{++} k^{--}, k^{++}\delta^{--}, \phi^0 \chi^0, \delta^0 \eta^0, \eta^0 \chi^0, \phi^0 \delta^0, \phi^0 \eta^0, \delta^0 \chi^0, \phi^+ \phi^-, \delta^+ \delta^-, \delta^{++} \delta^{--}, k^{++} k^{--}, \\
\frac{\eta^0 \eta^0}{\sqrt2}, \frac{\chi^0 \chi^0}{\sqrt2},
\frac{\phi^0 \phi^0}{\sqrt2}, \frac{\delta^0 \delta^0}{\sqrt2} \}$
leading to an 18$\times$18 matrix.
The 15 eigenvalues, analytically obtained are 
\bea
a_1 = a_2 = \l_1, \nonumber \\
a_3 = \l_1 + \l_2, \nonumber \\
a_4 = a_5 = \l_4 - 2 \l_5, \nonumber \\
a_6 = a_7 = \l_4 - \l_5, \nonumber \\
a_8 = a_9 = \l_4 + \l_5, \nonumber \\
a_{10} = \frac{1}{2} \Big(\l + \l_1 - \l_2 + \sqrt{(\l - \l_1 + \l_2)^2 + 16 \l_5^2} \Big), \nonumber \\
a_{11} = \frac{1}{2} \Big(\l + \l_1 - \l_2 - \sqrt{(\l - \l_1 + \l_2)^2 + 16 \l_5^2} \Big), \nonumber \\
a_{12} = a_{13} = \frac{1}{2} \Big(\l + \l_7 + \sqrt{(\l - \l_7)^2 + 8 \l_8^2} \Big), \nonumber \\
a_{14} = a_{15} = \frac{1}{2} \Big(\l + \l_7 - \sqrt{(\l - \l_7)^2 + 8 \l_8^2} \Big).
\eea

\textbf{Singly charged 2-particle states:}
A 12$\times$12 matrix is constructed in the basis 
$\{\delta^+ \phi^0, \delta^+ \eta^0, \delta^+ \delta^0, 
\delta^+ \chi^0, \phi^+ \phi^0, \phi^+ \eta^0, \phi^+ \delta^0, 
\phi^+ \chi^0, \delta^{++} \delta^-, \delta^{++} \phi^-,
k^{++} \delta^-, k^{++} \phi^- \}$. Its eigenvalues are
\bea
b_1 = a_1, ~
b_2 = a_3, ~
b_3 = a_4, ~
b_4 = a_6, ~
b_5 = b_6 = a_8, ~
b_7 = a_{10}, ~
b_8 = a_{11}, ~
b_9 = a_{12}, ~
b_{10} = a_{14}, \nonumber \\
b_{11} = \frac{1}{2} \Big(\l_4 + 2\l_5 + \l_6 + \sqrt{(\l_4 - 2\l_5 - \l_6)^2 + 24 \l_8^2} \Big), \nonumber \\
b_{12} = \frac{1}{2} \Big(\l_4 + 2\l_5 + \l_6 - \sqrt{(\l_4 - 2\l_5 - \l_6)^2 + 24 \l_8^2} \Big).
\eea

\textbf{Doubly charged 2-particle states:} We arrange the 2-particle states in the basis 
$\{\delta^{++} \phi^0, \delta^{++} \eta^0, \delta^{++} \delta^0, 
\delta^{++} \chi^0, k^{++} \phi^0, k^{++} \eta^0, k^{++} \delta^0, 
k^{++} \chi^0, \delta^+ \phi^+, \frac{\delta^+ \delta^+}{\sqrt2}, \frac{\phi^+ \phi^+}{\sqrt2} \}$. Amongst the total 11, 8 eigenvalues can be determined analytically as
\bea
c_1 = a_1, ~
c_2 = a_3, ~
c_3 = \frac{1}{2}(2 \l_1 + 3 \l_2), ~
c_4 = a_8, ~
c_5 = \l_6, ~
c_6 = \l_7, ~
c_7 = a_{12}, ~
c_8 = a_{14}.
\eea

\textbf{Triply charged 2-particle states:} A 4$\times$4 matrix is needed to be constructed in the basis, say, 
$\{\delta^{++}\delta^+, \delta^{++}\phi^+, k^{++}\delta^+, 
k^{++}\phi^+ \}$. The eigenvalues read
\bea
d_1 = a_1,~d_2 = a_6,~d_3 = c_5,~d_4 = c_6.
\eea

\textbf{Quadruply charged 2-particle states:} A 3$\times$3 matrix
constructed in the basis $\{\delta^{++}k^{++},
\frac{\delta^{++}\delta^{++}}{\sqrt2}, 
\frac{k^{++}k^{++}}{\sqrt2} \}$ has the following eigenvalues:
\bea
e_1 = a_1,~e_2 = c_6,~e_3 = \l_3.
\eea
The eigenvalues not determined analytically were computed numerically in the parameter space scans.

\subsection{Muon g-2 functions}
Analytical formulae for the integrals in $\Big(\Delta a_\mu\Big)_{\text{VLL}}$ are
\besub
\bea
I_1(m_1,m_2) &=& \Bigg[ m_1^6 - 6 m_1^4 m_2^2 + 3 m_1^2 m_2^4
+ 2 m_2^6 + 6 m_1^2 m_2^4 \text{log}\bigg(\frac{m_1^2}{m_2^2}\bigg) \Bigg]\bigg/ \big[6(m_1^2 - m_2^2)^4 \big], 
\\
I_2(m_1,m_2) &=& \Bigg[ m_1^4 - 4 m_1^2 m_2^2 + 3 m_2^4
+ 2 m_2^4\text{log}\bigg(\frac{m_1^2}{m_2^2}\bigg) \Bigg]\bigg/ \big[2(m_1^2 - m_2^2)^3 \big], \\
I_3(m_1,m_2) &=& \Bigg[ 2 m_1^6 + 3 m_1^4 m_2^2 - 6 m_1^2 m_2^4
+ m_2^6 - 6 m_1^4 m_2^2 \text{log}\bigg(\frac{m_1^2}{m_2^2}\bigg) \Bigg]\bigg/ \big[6(m_1^2 - m_2^2)^4 \big], \\
I_4(m_1,m_2) &=& \Bigg[ m_1^4 - m_2^4
- 2 m_1^2 m_2^2\text{log}\bigg(\frac{m_1^2}{m_2^2}\bigg) \Bigg]\bigg/ \big[2(m_1^2 - m_2^2)^3 \big].
\eea
\eesub

\subsection{One-loop beta functions}
The one-loop beta function for a quartic coupling $\l_i$ is split into scalar, gauge and fermionic terms as
$\beta_{\l_i} = \beta_{\l_i}^S + \beta_{\l_i}^g + \beta_{\l_i}^F$. Thus,
\besub
\bea
16 \pi^2 \beta^S_{\l} &=& 12 \l^2 + 6 \l_4^2 + 4 \l_5^2
+ 2 \l_6^2 + 4 \l_8^2, \\ 
16 \pi^2 \beta^S_{\l_1} &=& 14 \l_1^2 + 4 \l_1 \l_2 + 2 \l_2^2 + 4 \l_4^2 + 4 \l_5^2 + 2 \l_7^2, \\
16 \pi^2 \beta^S_{\l_2} &=& 12 \l_1 \l_2 + 3 \l_2^2 - 8 \l_5^2, \\
16 \pi^2 \beta^S_{\l_3} &=& 10 \l_3^2 + 4 \l_6^2 + 6 \l_7^2, \\
16 \pi^2 \beta^S_{\l_4} &=& 6 \l \l_4 + 8 \l_1 \l_4 
+ 2 \l_2 \l_4 + 4 \l_4^2 + 8 \l_5^2 + 2 \l_6 \l_7 + 4 \l_8^2, \\
16 \pi^2 \beta^S_{\l_5} &=& 2 \l \l_5 + 2 \l_1 \l_5
- 2\l_2 \l_5 + 8 \l_4 \l_5 - 4 \l_8^2, \\
16 \pi^2 \beta^S_{\l_6} &=& 6 \l \l_6 + 4 \l_3 \l_6
+ 4 \l_6^2 + 6 \l_4 \l_7 + 12 \l_8^2, \\
16 \pi^2 \beta^S_{\l_7} &=& 4 \l_4 \l_6 + 8 \l_1 \l_7
+ 2 \l_2 \l_7 + 4 \l_3 \l_7 + 4 \l_7^2 + 4 \l_8^2, \\
16 \pi^2 \beta^S_{\l_8} &=& 2 \l \l_8 + 4 \l_4 \l_8
- 8 \l_5 \l_8 + 4 \l_6 \l_8 + 2 \l_7 \l_8.
\eea
\eesub

\besub
\bea
16 \pi^2 \beta^g_{\l} &=& - 3\l(g_1^2 + 3 g_2^2) 
+ \frac{3}{4}g_1^4 + \frac{3}{4}g_1^2 g_2^2 
+ \frac{9}{4}g_2^4, \\
16 \pi^2 \beta^g_{\l_1} &=& - 12\l_1(g_1^2 + 2 g_2^2) 
+ 12 g_1^4 + 24 g_1^2 g_2^2 + 18 g_2^4,  \\
16 \pi^2 \beta^g_{\l_2} &=& - 12\l_2(g_1^2 + 2 g_2^2) 
- 48 g_1^2 g_2^2 
+ 12 g_2^4, \\
16 \pi^2 \beta^g_{\l_3} &=& - 48\l_3 g_1^2 + 192 g_1^4, \\
16 \pi^2 \beta^g_{\l_4} &=& - \l_4 
\Big( \frac{15}{2} g_1^2 + \frac{33}{2} g_2^2 \Big)
+ 3 g_1^4 + 6 g_2^4, \\
16 \pi^2 \beta^g_{\l_5} &=& - \l_5 
\Big( \frac{15}{2} g_1^2 + \frac{33}{2} g_2^2 \Big)
- 6 g_1^2 g_2^2, \\
16 \pi^2 \beta^g_{\l_6} &=& - \l_6\Big(\frac{51}{2} g_1^2 + 9 g_2^2 \Big) + 12 g_1^4, \\
16 \pi^2 \beta^g_{\l_7} &=& - \l_7\Big(20 g_1^2 + 8 g_2^2 \Big) + 48 g_1^4, \\
16 \pi^2 \beta^g_{\l_7} &=& - \l_8\Big(11 g_1^2 + 7 g_2^2 \Big) + 48 g_1^4.
\eea
\eesub

\besub
\bea
16 \pi^2 \beta^F_{\l} &=& 4\l \Big(3 y_t^2 + 3 y_b^2 +  y_\tau^2 +  y_4^2
+ (y_4^\prime)^2 \Big)
- 4 \Big(3 y_t^4 + 3 y_b^4 + y_\tau^4 + y_4^4
+ (y_4^\prime)^4 \Big), \\
16 \pi^2 \beta^F_{\l_1} &=& 16 \l_1 
\Big(y_\Delta^{\mu 4} \Big)^2 - 64 \Big(y_\Delta^{\mu 4} \Big)^4, \\
16 \pi^2 \beta^F_{\l_2} &=& 16 \l_2 
\Big(y_\Delta^{\mu 4} \Big)^2 + 64 \Big(y_\Delta^{\mu 4} \Big)^4, \\
16 \pi^2 \beta^F_{\l_3} &=& 16 \l_3 
\Big(y_S^{\mu 4} \Big)^2 - 64 \Big(y_S^{\mu 4} \Big)^4, \\
16 \pi^2 \beta^F_{\l_4} &=& \l_4 
\bigg[ \Big(8 y_\Delta^{\mu 4} \Big)^2 + 6 y_t^2
+ 6 y_b^2 + 2 y^2_\tau + 2 y_4^2 + 2 (y_4^\prime)^2 \bigg], \\
16 \pi^2 \beta^F_{\l_5} &=& \l_5 
\bigg[ \Big(8 y_\Delta^{\mu 4} \Big)^2 + 6 y_t^2
+ 6 y_b^2 + 2 y^2_\tau + 2 y_4^2 + 2 (y_4^\prime)^2 \bigg], \\
16 \pi^2 \beta^F_{\l_6} &=& \l_6 
\bigg[ \Big(8 y_\Delta^{\mu 4} \Big)^2 + 6 y_t^2
+ 6 y_b^2 + 2 y^2_\tau + 2 y_4^2 + 2 (y_4^\prime)^2 \bigg], \\
16 \pi^2 \beta^F_{\l_7} &=& \l_7 
\bigg[\Big(8 y_\Delta^{\mu 4} \Big)^2 + \Big(8 y_S^{\mu 4} \Big)^2 + 16 y_4^2 \Big(y_\Delta^{\mu 4} \Big)^2  \bigg], \\
16 \pi^2 \beta^F_{\l_8} &=& \l_8 
\bigg[\Big(4 y_\Delta^{\mu 4} \Big)^2 + \Big(4 y_S^{\mu 4} \Big)^2 + 6 y^2_t + 6 y^2_b + 2 y^2_\tau + 2 y_4^2 
+ (y_4^\prime)^2 \bigg].
\eea
\eesub

We next list the $\beta$-functions for the relevant Yukawa couplings below.
\besub
\bea
16 \pi^2 \beta_{y_t} &=& \frac{9}{2}y^3_t + y_t
\big(3 y_b^2 + y_\tau^2 + y^2_4 + (y^\prime_4)^2 - \frac{17}{12} g_1^2 - \frac{9}{4} g_2^2 - 8 g^2_3 \big),
\\
16 \pi^2 \beta_{y_b} &=& \frac{9}{2}y^3_b + y_b
\big(3 y_t^2 + y_\tau^2 + y^2_4 + (y^\prime_4)^2 - \frac{5}{12} g_1^2 - \frac{9}{4} g_2^2 - 8 g^2_3 \big), \\
16 \pi^2 \beta_{y_\tau} &=& \frac{5}{2}y^3_\tau
+ y_\tau
\big(3 y_t^2 + 3 y_b^2 + y^2_4 + (y^\prime_4)^2 - \frac{15}{4} g_1^2 - \frac{9}{4} g_2^2 \big), \\
16 \pi^2 \beta_{y_4} &=& \frac{5}{2}y^3_4
+ y_4
\big(3 y_t^2 + 3 y_b^2 + (y^\prime_4)^2
 - \frac{15}{4} g_1^2 - \frac{9}{4} g_2^2 \big), \\
16 \pi^2 \beta_{y_4^\prime} &=& \frac{5}{2}
(y_4^\prime)^3
+ y_4^\prime
\big(3 y_t^2 + 3 y_b^2 + y^2_4
 - \frac{15}{4} g_1^2 - \frac{9}{4} g_2^2 \big), \\
16 \pi^2 \beta_{{y^{\mu 4}_\Delta}} &=& 
8 (y_\Delta^{\mu 4})^3 + y_\Delta^{\mu 4} 
\big( \frac{y_4^2}{2} - \frac{3}{2} g_1^2
- \frac{9}{2} g_1^2 \big), \\
16 \pi^2 \beta_{{y^{\mu 4}_S}} &=& 
8 (y_S^{\mu 4})^3 + y_S^{\mu 4} 
\big( y_4^2 - 6 g_1^2 \big).
\eea
\eesub

Finally, the $\beta$-functions for the gauge couplings read
\besub
\bea
16 \pi^2 \beta_{g_1} &=& \frac{67}{6} g_1^3, \\
16 \pi^2 \beta_{g_2} &=& -\frac{13}{6} g_2^3, \\
16 \pi^2 \beta_{g_3} &=& -7 g_3^3.
\eea
\eesub

\bibliography{ref} 
\end{document}